\documentclass[english,12pt]{article}
\usepackage[utf8]{inputenc}
\usepackage[T1]{fontenc}
\usepackage[numbers,sort&compress]{natbib}
\usepackage{afterpage}
\usepackage{amsmath}
\usepackage{amsfonts} 
\usepackage{caption}
\usepackage{subcaption}
\usepackage{longtable}
\usepackage{pdflscape}
\usepackage{adjustbox}
\usepackage[toc,page]{appendix}
\usepackage{amssymb}
\usepackage{comment}
\usepackage[shortlabels]{enumitem}
\usepackage{gensymb}
\usepackage{siunitx}
\usepackage{graphicx}
\usepackage{float}
\usepackage{bm}
\usepackage{multicol}
\usepackage{mathtools}
\usepackage{multirow}
 \usepackage{booktabs}
\usepackage{lscape} 
\usepackage{csquotes}
\usepackage{setspace}
\usepackage[makeroom]{cancel}
\usepackage[table,xcdraw]{xcolor}
\usepackage{array}
\newcolumntype{P}[1]{>{\centering\arraybackslash}p{#1}}

\usepackage{etoolbox}
\AtBeginEnvironment{quote}{\par\singlespacing\small}
\usepackage{sansmath}
\usepackage{wrapfig}
\graphicspath{ {figures/} }
\usepackage{array}
\usepackage[paper=a4paper,top=25mm,left=25mm,right=25mm,bottom=25mm]{geometry}
\usepackage{fancyhdr}
\usepackage[pangram]{blindtext} 
\usepackage[font=small]{caption} 
\usepackage[colorlinks=true, urlcolor=blue,  linkcolor=blue, citecolor=blue]{hyperref} 

\usepackage{xcolor} 
\makeatletter 
\renewcommand{\fnum@figure}{\textbf{\figurename~\thefigure}}
\makeatother
\makeatletter 
\renewcommand{\fnum@table}{\textbf{Table~\thetable}}
\makeatother

\title{Geographic distribution of the global agricultural workforce every decade for the years 2000-2100}

\author{Naia Ormaza Zulueta \footnote{Department of Environmental Studies, University of Colorado, Boulder, CO, USA\\
\indent\, $^\dagger$ Better Planet Lab, University of Colorado, Boulder, CO, USA \\
\indent\, $^\ddagger$ Mortenson Center for Global Engineering and Resilience, University of Colorado, Boulder, CO, USA \\
\indent\, contact: Naia.OrmazaZulueta@colorado.edu,
Zia.Mehrabi@colorado.edu}\,\,$^\dagger$\\
Steve Miller$^\ast$\\
Zia Mehrabi$^\ast$$^\dagger$$^\ddagger$}

\date{\today}

\begin{document}
\maketitle
\begin{abstract}
Agricultural workers play a vital role in the global economy and food security by cultivating, transporting, and processing food for populations worldwide. Despite their importance, detailed spatial data on the global agricultural workforce have remained scarce. Here, we present a new gridded dataset that maps the global distribution of agricultural workers for every decade over the years 2000-2100, distributed at 0.083$\times$0.083 degrees resolution, roughly $\sim$10km$\times$10km at the Equator. The dataset is developed using an empirical modeling framework relying on generalized additive mixed models (GAMMs) that integrate socioeconomic variables, including gross domestic product per capita, total population, rural population size, and agricultural land use. The predictions are consistent with Shared Socio-economic Pathways and we distribute full time series data for all SSPs 1 to 5. This dataset opens new avenues for future research on labour force health, productivity and risk, and could be very useful for developing informed, forward-looking strategies that address the challenges of climate resilience in agriculture. The dataset and code for reproducing it are available for the user community [publicly available on publication at DOI: \href{https://doi.org/10.5281/zenodo.14443333}{10.5281/zenodo.14443333}].
 
\end{abstract}

\newgeometry{top=35mm,bottom=25mm,right=25mm,left=25mm}

\section{Introduction}
Understanding the dynamics of the global agricultural workforce, both currently and in the future, is crucial for advancing key development objectives such as labour efficiency, global food security, and worker well-being. This need has become even more pressing, for example, in the context of mobility constraints caused by events like the COVID-19 pandemic, when restrictions on movement severely disrupted agricultural production and labour markets \cite{IPES-Food2024, barret2020actions}. Likewise, climate change poses a significant threat, with rising temperatures in regions like South Asia already leading to increased cases of heat exhaustion among agricultural workers, reducing productivity, threatening livelihoods, and leading to migration. This phenomenon is not limited to a single region; similar impacts have been observed globally, across countries from all regions (e.g., Spain, USA, Indonesia, Nigeria), and are starting to become a frequent phenomenon \cite{khayat_2022_impacts, de_Lima_2021, nelson2024global}.

\bigskip

Despite the critical importance of a high-resolution representation of agricultural workforce data for proactive and reactive response, such data does not exist in the public domain. Many studies have quantified the effect of climate change or other shocks on crop yields, but far less is known about the impact of these shocks on the workforce. While some efforts have been made to examine larger structural changes, such as labour reallocation in developing economies due to rising temperatures (e.g., Liu et al. (2023) \cite{liu2023climate}) as well as impacts of heat stress on agricultural workers (e.g., Nelson et al. (2024) \cite{nelson2024global}), these efforts rely on coarse workforce data, and use current distributions of the percentage employed population in future estimates. This approach misses not only important variations in exposure that occur at higher spatial resolutions, but also how the workforce may change in the future. Such data are essential for understanding subnational trajectories within countries and how different hazards, including extreme weather, pesticides, pollution, and epidemics interact with different exposures in the labour force.

\bigskip

To fill this gap, we present a new dataset detailing the distribution of the global agricultural workforce in contemporary times and in the future (every decade from 2000-2100). The data we present are based on a theoretically grounded empirical framework recently developed to predict the number and size of farms from socioeconomic predictors \cite{mehrabi2023likely}, and are consistent with Shared Socio-economic Pathways (SSP). Here, we adapt this prior work to model the proportion of people working in agriculture around the globe, adjusting the framework to enable spatiotemporal predictions (for example, including features that represent economic attractors, like urban centers, which pull population into non-agricultural sectors). Given prior theory, the basic expectation is that increases in rural population will increase the agricultural workforce, up to a certain limit determined by the availability of agricultural lands, but that this relationship is mediated by regional economic factors which represent rural-urban migration and non-agricultural sector growth \cite{mehrabi2023likely, eastwood2010farm, jayne2014land, taylor2001migration}. Here we focus on SSP2 -- outputs for all SSP scenarios are available for researchers and policymakers interested in exploring alternative future pathways in the full dataset. 

\bigskip

Previous studies have employed machine learning, particularly tree-based approaches, for downscaling various socioeconomic and environmental variables, such as wealth \cite{chi_microestimates_2022}, poverty \cite{lee_high-resolution_2022}, water and sanitation \cite{deshpande_mapping_2020}, food security indicators \cite{osgood-zimmerman_mapping_2018}, and agricultural lands \cite{mehrabi2024globalagri}. Here we rely on Generalized Additive Mixed Models (GAMMs) because they offer several out-of-the-box advantages for fitting flexible functional forms,  interpretation, and non-naive extrapolation. We do note that the majority of downscaling exercises, such as those aforementioned, have limited validation, and rarely possess labels on the scale of the final deploy. Here we present an advance that attempts to compute this uncertainty, and our validation pipeline computes spatial, temporal and scale dependent prediction error.

\bigskip

The data we present not only captures a critical component of the global labour force and that of the food systems that sustain us, but provides an opportunity to better understand the risks facing this workforce, labour efficiency, climate resilience, socioeconomic shocks, and more. For example, integrating workforce projections with climate projection data might help better assess future climate impacts on agricultural productivity and worker health \cite{nelson2024global} or provide critical insights into migration patterns \cite{mueller_heatstress_2014}. It could also provide insights into labour market demands through potential reductions in working hours due to high levels of heat exposure, or workers' health, for example by analysing the exposure to pesticides and pollution (e.g. particulate matter, nitrogen oxide, ozone), whether from farm activities, wildfires or industrial sources. As such we are excited to release these data to the research community to advance work and use in downstream analyses.

\section{Methods}
\subsection{Overview}
\begin{figure}[ht]
    \centering
    \includegraphics[width=\textwidth]{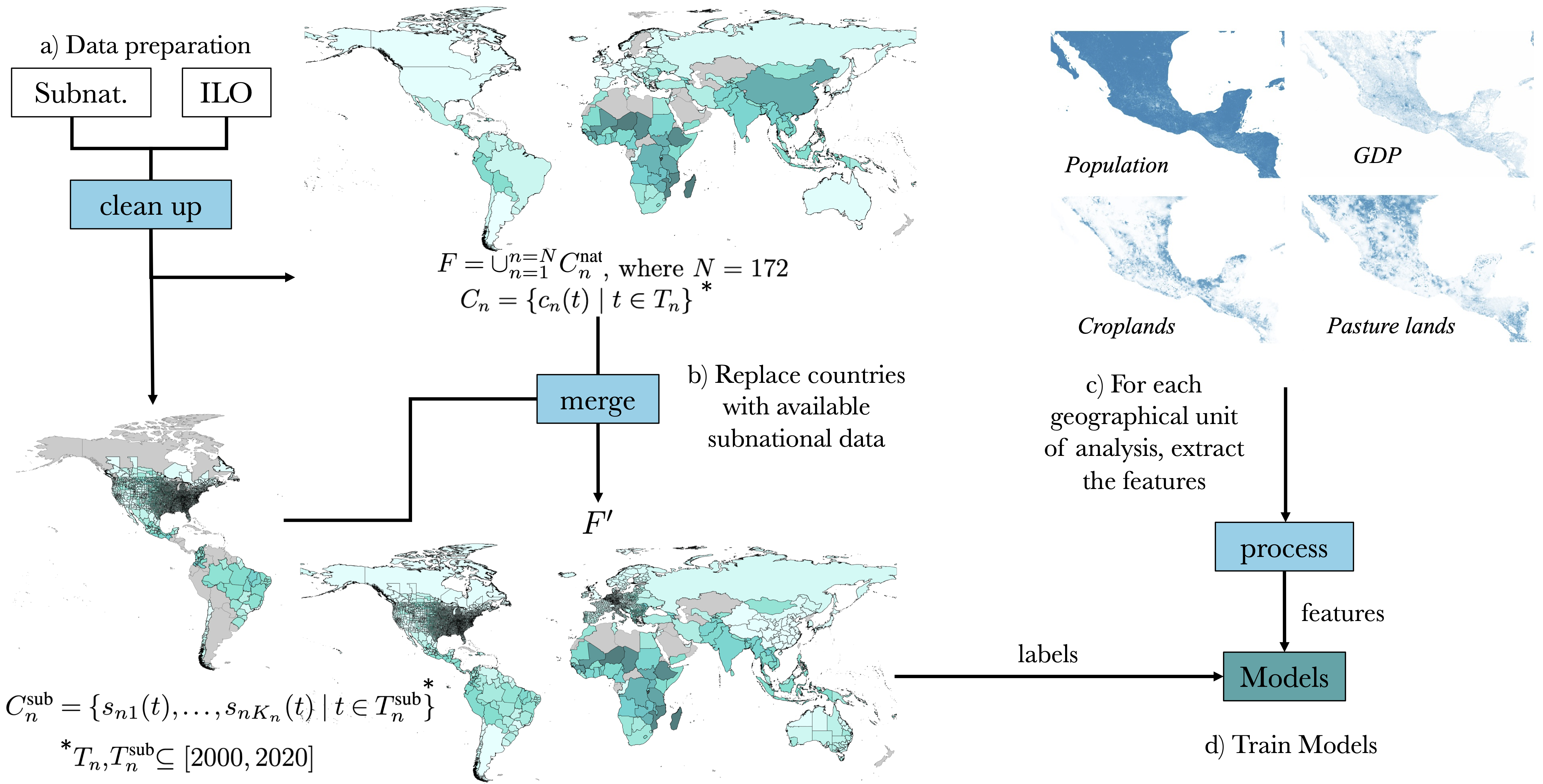}
    \caption{Data pre-processing and training pipeline.}
    \label{fig:main_panel1}
\end{figure}

\begin{figure}[ht]
    \centering
    \includegraphics[width=\textwidth]{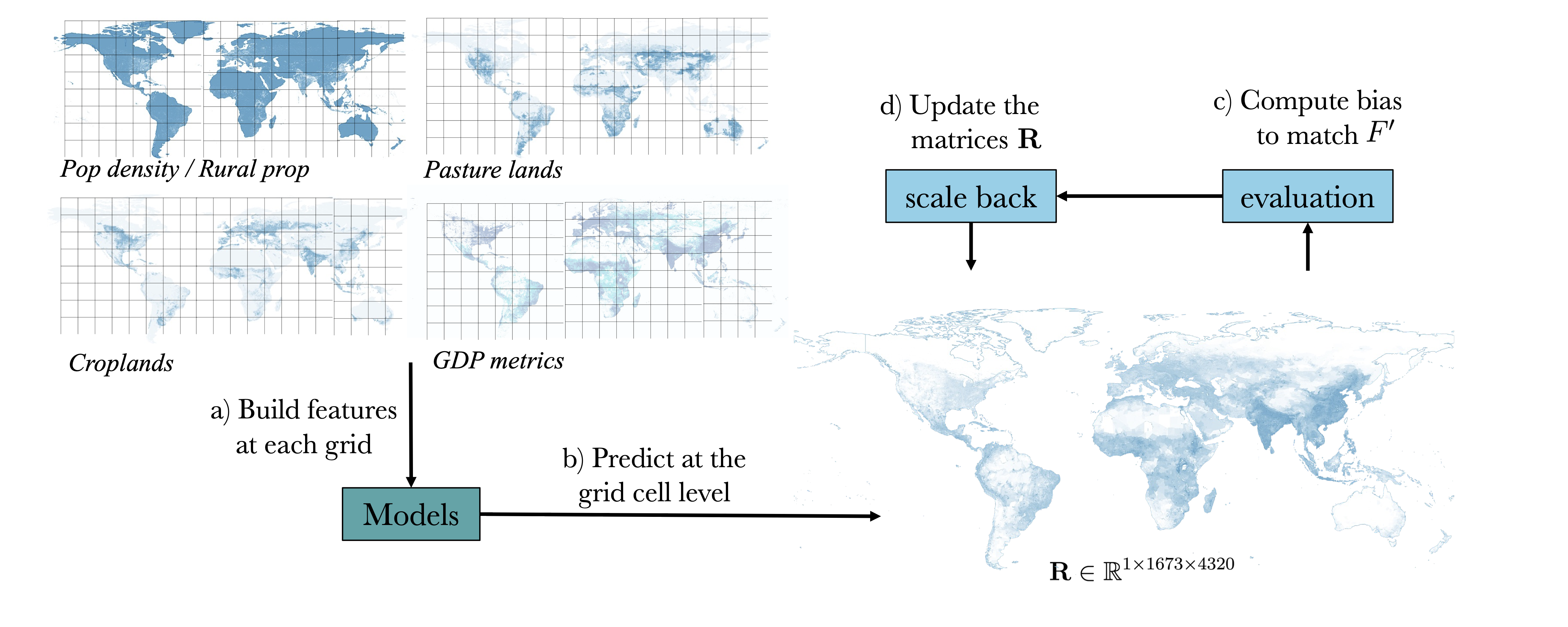}
    \caption{Evaluation and post-processing.}
    \label{fig:main_panel2}
\end{figure}

The data development and analysis pipelines are set up to model and predict the proportion of the Employed Population Working in Agriculture (EPWA) in a geographic unit as a function of several socioeconomic features. The data development pipeline,  summarized in Figure \ref{fig:main_panel1}, trains GAMMs after cleaning, harmonizing, and merging features and the target. The analysis pipeline performs comparable preprocessing steps on features, predicts over these using the trained model, and optionally bias-corrects predictions, as summarized in Figure \ref{fig:main_panel2}.  Details of these pipelines are provided below.
\subsection{Input data}

\subsubsection{Labels}

We first compiled national EPWA for all available country-year pairs during 2000-2020 from the International Labor Organization (ILO). The ILO data contains a total of 172 country-level observations. For the same time period, we also compiled all available subnational EPWA data from several countries – Canada, 30 countries in the EU, Australia, the United States, Mexico, Colombia, Brazil, Chile. The subnational dataset comprises approximately 90 thousand records spanning 4,560 geographic units (Administrative level 1-2) from a total of 37 countries spanning a wide range of income and development groups. We use the following definition of agricultural worker: ``all persons of working age who carried out agricultural work (activities in agriculture, hunting, forestry and fishing)''. We do note that the data for Australia and EUROSTAT consider agriculture, forestry and fishing (no hunting). The working age can vary spatially; the subnational statistics we collected only consider all persons above 18, whereas countries reporting in the  ILO national statistics can include youth and adults 15+. Finally, we also know that agricultural sectors often rely on international migrant workers, and these and the undocumented workforce are not included in our data. See Appendix \ref{sec:labels} for detailed information on the availability of national and subnational data across sources. 

\bigskip

Our final set of training labels combines the subnational and national EPWA records, using national data only if subnational data does not exist for a particular country. Formally, denote this set of country-level EPWA records as \( F = \cup_{n=1}^{n=N} C^{\text{nat}}_n \), where \( N = 172 \). Each element \( C^{\text{nat}}_n \) in the set \( F \) represents a set of country-level observations, which may have data available for different years \( t \in [2000, 2020] \). That is, \( C^{\text{nat}}_n = \{c_n(t) \mid t \in T^{\text{nat}}_n\} \), where \( T^{\text{nat}}_n \subseteq [2000, 2020] \) denotes the years for which national data is available for country \( n \), and \(c_n(t)\) is a single observation at the country-year level. In addition to these national records, some countries also have multiple subnational level observations; denote that set of countries by \(S\), with \(N^{\text{sub}}=|S|=37\). For a country \( n \) with \( K_n \) subnational units (e.g., divisions, counties, provinces, etc.), \( C^{\text{sub}}_n = \{s_{n1}(t), \ldots, s_{nK_n}(t) \mid t \in T^{\text{sub}}_n\} \), where \( T^{\text{sub}}_n \subseteq [2000, 2020] \) represents the years for which data is available for the subnational units of country \( n \). To create a new set \( F' \) that combines both national and subnational observations, we define: 

\begin{equation}
    F' =  \bigcup_{n \in \{N\setminus S\}} C^{\text{nat}}_n \cup \bigcup_{n \in S} C_n^{\text{sub}}
\end{equation}

\subsubsection{Features}
We use three input datasets to construct features for the analysis, selected based on prior empirical work \cite{mehrabi2023likely}. The first is the proportion of land areas used as cropland (described as ``arable land'', ``land in crops'', ``fallow land'', ``cultivated land'' and ``temporary meadows'') and pasture lands (``permanent meadows'', ``grazing
land'', ``pasture land'') in the year 2000 at $0.08\overline{3}$ x $0.08\overline{3}$-degree resolution, developed by Ramankutty et al. (2000) \cite{ramankutty_farming_2008}.  The second is rural and urban population size, from Gao (2017) \cite{gao2017downscaling}, $0.008\overline{3}$ x $0.008\overline{3}$ degrees downscaled from the one-eighth degree data published in Jones and O'Neill (2016) \cite{Jones_2016}. Third, we proxy economic output with the median of the gross domestic product per capita (GDP per capita) distribution (to proxy the economic opportunity of an area) within a region using gridded GDP per capita from the $0.008\overline{3}$ x $0.008\overline{3}$ degrees by Wang and Sun (2022) \cite{wang_global_2022}.

\bigskip

To convert the gridded datasets (rural proportion, total population, croplands and pasture land areas, and GDP per capita) into trainable format, we compute zonal statistics for each geographic unit. The derived features are, therefore, the ratio of rural population to total population, total population density, cropland and pastureland mean fractional areas, and median GDP per capita for each administrative unit. For GDP per capita features were available on a annual basis from \cite{wang_global_2022}, and for population we interpolate between the decadal time series to obtain annual estimates to match to labels (see Appendix \ref{popinterpolate} for more information). We keep cropland and pasture constant in all training and future projections. It is important to note the discrepancies between land/sea masks for population and GDP layers compared to cropland and pasture files, resulting in omitted coastal grid cells. This mainly affects urban, densely populated areas, likely having minimal impact (more significant in smaller countries or islands) on agricultural worker estimates. 

\subsection{Modeling approach}

We rely on prior work and the theoretical background on agricultural sector growth to define the inputs into our model \cite{mehrabi2023likely, eastwood2010farm, jayne2014land, taylor2001migration}, but explore various model structures and specifications. On structure, we expand on previous work to include interaction terms in the model to allow for the relationship between GDP per capita and agricultural workforce to depend on population density and rural proportions. This allows for the possibility that the economic pull of urban areas may be weaker in regions with lower rural densities as opposed to higher rural densities, which may be influenced more strongly by an economic attractor. Likewise, in wealthier nations, peri-urban agriculture where farms serve nearby cities might sustain a higher proportion of the workforce, especially in the context of higher income levels \cite{zasada_multifunctional_2011}. Adding interactions allows for these nuances and the possibility of the model capturing the added complexities of local economic and demographic conditions that influence agricultural employment dynamics (see Figure \ref{fig:residual_relationships}).

\bigskip

The primary specification is a Beta regression GAMM with a logit link that considers the ratio of the agricultural workforce in a geographical unit for any given year as a function of the logarithm of the unit's median GDP per capita, ($\ln(\text{G}_{50})$), agricultural lands ($\ln(\text{AL})$), rural proportion ($\ln(\text{R})$), and population density ($\ln(\text{P})$). We incorporated random country-specific variations to account for the correlation of errors and non-independence of data collected from the same country over multiple years. The generalized additive model (Eq. \ref{eq:model_gam}) was parameterized, generating estimates of \( \mu_i \), which is \( \mathbb{E}(y) \), of the agricultural workforce ratio in a given geographic unit in a given year. We choose the Beta distribution for convenience of fixing the predictions between zero and one. 

\begin{equation}
    y_i \sim \text{Beta}(\alpha_i, \beta_i) 
    \quad \text{where} \quad \alpha_i = \mu_i \phi \quad \text{and} \quad \beta_i = (1 - \mu_i) \phi
    \label{eq:beta_distribution}
\end{equation}

\begin{equation}
    \mu_i = \text{logit}^{-1} \Bigg( \beta_0  
     + \sum_k^K \sum_{h_k=1}^{H_k} \gamma_{h_k} b_{h_k}(X_{ik}) 
     + \sum_{l=1}^2 \sum_{m_l=1}^{M_l} \gamma_{m_l} b_{m_l}\left(\ln(\text{G}_{50i}), X_{il})\right)  +  \delta_{j[i]} \Bigg)
    \label{eq:model_gam}
\end{equation}

\begin{equation}
    \text{for }i=1, \dots N\text{ observations, and } 
X = [\ln(\text{R}), \ln(\text{P}), \ln(\text{G}_{50}), \ln(\text{AL})]
\end{equation}

Where \( y_i \) is the observed proportion of the agricultural workforce, \( \phi \) is a precision parameter, \( \beta_0 \) is the intercept, \(\gamma_{h_k}\) and \(\gamma_{h_l}\) are the spline coefficients; \(b_{h_k}\left(\cdot\right)\) is the thin-plate spline basis function for $X$, for the \(i\)th observation (with \(H_k\) basis functions); \(b_{m_l}\left(\cdot\right)\) are the spline basis functions for the interaction terms; and \(\delta_{j[i]}\) is a random effect term for country-level variation for countries $j=1, \dots J$.  We also fit an equivalent model with \(\delta_{j[i]}\) where $j=1, \dots J$ represent UN geographical subregions.  We also explore a simple linear model, qGAM models with smooth terms with and without random effects, and a range of tree-based regression models, alongside the problems they pose for this task (see Appendix \ref{sec:additional_exp}).

\subsection{Validation Tests}

To address the problem of lack of validation at the deployment scale, which plagues many socioeconomic down scaling exercises, we implement a multi-stage validation procedure which includes temporal, spatial and multi-scale validation. These strategies allow us to assess the model's performance across different geographies and time periods. In our analysis, we assume that the relationships captured by the model in the baseline years remain consistent over time. Similarly, we assume that the spatial relationships identified at higher scales are maintained at lower scales. The validity of these assumption are investigated through temporal and multi-scale spatial validation tests. 

\bigskip

In the spatial validation strategy, we evaluate the model's ability to predict agricultural workforce ratios in different subnational units from what the model is trained on. We use random stratified 80-20 train test split sampling from subnational units of all $C_m^{\text{sub}}$ countries, and evaluate the ability of the model to fill gaps.

\bigskip

For multi-scale validation, we aim to explore how much error arises from predicting on finer resolution features than the model is trained on. We define: \( F''^{(i)}_{\text{train, scale}} \) as the training dataset for region \( R_i \), which includes the entire global dataset \( F' \) except for the subnational units of countries with subnational data within \( R_i \), replaced by their aggregated national data; and \( F''^{(i)}_{\text{valid, scale}} \) as the validation dataset to simply be all the $C_m^{\text{sub}}$ countries in $R_i$. To formally describe the datasets:

\begin{equation}
    F''^{(i)}_{\text{train, scale}} = \left( F' \setminus \{C_n^{\text{sub}} \mid n \in R_i\} \right) \cup \bigcup_{n\in R_i} C_n^{\text{nat}}
\end{equation}

and

\begin{equation}
    F''^{(i)}_{\text{valid, scale}} = \{C_n^{\text{sub}} \mid n\in R_i\}
\end{equation}

Models are then trained on \( F''^{(i)}_{\text{train, scale}} \) and evaluated on \( F''^{(i)}_{\text{valid, scale}} \) for each region \( R_i \) where \( R = \{R_1, \ldots, R_{22}\} \) is the set of 22 UN geoscheme regions.  

\bigskip

In the temporal validation strategy, we assess the model's ability to forward-predict and back-predict. We divide $F'$ into two temporal subsets. First, for forward-predicting, we train on the first 80\% of the data (2000-2017) and validate on the last 20\% (2018 onward). Let \( T = [2000, 2020] \)\footnote{The temporal subsets are built upon data availability, which differs across years. See table \ref{tab:national_labels_1} for more information.} represent the entire time period. We define \( T_{\text{train,forward}} = [2000, 2017] \) as the training period for forward-predicting; and, \( T_{\text{valid,forward}} = [2018, 2020] \) as the validation period for forward-predicting. Thus, the training and validation datasets for forward-predicting are:

\begin{equation}
F'_{\text{train,forward}} = \{C_n(t) \mid C_n \in F', t \in [2000, 2017]\}
\end{equation}

\begin{equation}
F'_{\text{valid,forward}} = \{C_n(t) \mid C_n \in F', t \in [2018, 2020]\}
\end{equation}

For back-predicting, we simply switch the years used to train and test, training on future data and predicting in the past:

\begin{equation}
F'_{\text{train,backward}} = \{C_n(t) \mid C_n \in F', t \in [2005, 2020]\}
\end{equation}

\begin{equation}
F'_{\text{valid,backward}} = \{C_n(t) \mid C_n \in F', t \in [2000, 2004]\}
\end{equation}

\subsection{Deployment}

To deploy the model, we use the same set of predictors for every decade between years 2000 - 2100 under all SSPs (SSP1 to SSP5). All rasters in the stack are resampled to a common extent $(-180\degree, 180\degree, -56\degree, 84\degree)$ and resolution ($0.083\times 0.083$ degrees). Next, we compute the median GDP per capita at administrative unit level 2  and allocate that to every grid cell in that unit (we do not include local income in deployment at the grid cell). All other grid cell level data for variables are such as rural proportions, population density and cropland and pasture fractional area are kept as features, alongside geographical context (country and region). The regional model is used when there is no matching country in the training set, allowing us to predict for grid cells in countries not included in the training data. For example, Kazakhstan and Turkmenistan are two of the countries missing from our training data. By using regional-level random effects, we apply the random effect from a subregion including Kyrgyzstan, Tajikistan, and Uzbekistan. We do note we compute the mean instead of the sum of the agricultural lands because we want to maintain scale-free for deployment of the model, but maintain constant agricultural area in future predictions.

\subsection{Post-processing}
\subsubsection{Scaling back}
A straightforward approach to correcting the error in prediction (which we optionally do for maintaining consistency with original statistics in the observational period), is to scale back the error to the grid cells. Let \( EPWA_{i,t}^{\textbf{expected}} \) represent the true ratio of the agricultural workforce for unit \( i \)  in year \( t \), and \( EPWA_{i,t}^{\textbf{predicted}} \) be the predicted ratio for unit \( i \) in year \( t \). The predicted agricultural population for unit \( i \) in year \( t \) is \( A_{i,t} = EPWA_{i,t}^{\textbf{predicted}} \times N_{i,t} \times R_{i,t}\), where \( N_{i,t} \) is the total population and \( R_{i,t} \) is the employable-to-total ratio for unit \( i \) in year \( t \). The aggregated agricultural population for unit \( i \) is \( \sum_{j\in G_i} A_{j} \), and the aggregated total population is \( \sum_{j\in G_i} N_{j} \), $\forall j \in G_i $, where $G_i$ represents the set of all grid cells within unit $i$. We assume the employable-to-population ratio to be the same for any grid cell in a unit $i$ time $t$. 

\bigskip

We define the correction factor for unit \( i \) in year \( t \) as 

\begin{equation}
    \xi_{i,t} = \frac{\sum_{j\in G_i} EPWA_{j,t}^{\textbf{expected}}\cdot N_{j,t}}{\sum_{j\in G_i} EPWA_{j,t}^{\textbf{predicted}}\cdot N_{j,t}}
\end{equation}

We apply the correction factor to all predicted ratios in unit $i$ and time $t$ to compute the corrected ratio. While this strategy can assist to remove bias to some extent, it can also result in severe boundary disparities between administrative units. Substantial variations in predictions at the boundaries are reasonable given that agricultural labour shares can vary significantly across administrative units due to changes in laws, regulations, or cultural contexts and so we leave these in place, and so we do not smooth the version of the dataset which is bias corrected. Because we understand that some users may prefer to use the bias corrected data for alignment with the data, we provide both the corrected and the uncorrected maps. Corrected estimates are distributed for users wanting consistency with ILO data.

\section{Results and Discussion}
\subsection{Model}
\label{sec:model_comparison}
\subsubsection{Model structure exploration}
The comparison of different model structures outlined in Table \ref{tab:model_comparison_gam}  indicates how adding smooth terms for the predictors significantly improves performance showcasing the importance of capturing non-linear relationships and country-specific variations. There is a slight improvement on the performance with the introduction of interaction terms. We move with the best fitting model from the set for deployment.

\begin{table}[!h]
\centering
\caption{Performance metrics for different GAM models, including Generalized Cross-Validation (GCV), Akaike Information Criterion (AIC), Explained Variance, R-Squared and RMSE.}
\centering
\begin{tabular}[t]{lrrrr}
\toprule
\textbf{Model} & \textbf{GCV} & \textbf{AIC} & \textbf{Expl. Var.} & \textbf{R}$^2$\\
\midrule
Linear model & -168068.6 & -336170.8 & 0.58 & 0.36\\
Smooth terms & -172601.7 & -345417.1 & 0.63 & 0.43\\
Smooth terms + RE & -201015.3 & -402808.9 & 0.83 & 0.74\\
Smooth terms + RE + Int & -203359.1 & -407662.0 & 0.84 & 0.76\\
\bottomrule
\end{tabular}
\label{tab:model_comparison_gam}
\end{table}

\subsubsection{Visual fits for selected model}
\begin{figure}[ht]
    \centering
    
    \begin{subfigure}[b]{0.49\textwidth}
        \centering
        \includegraphics[width=\textwidth]{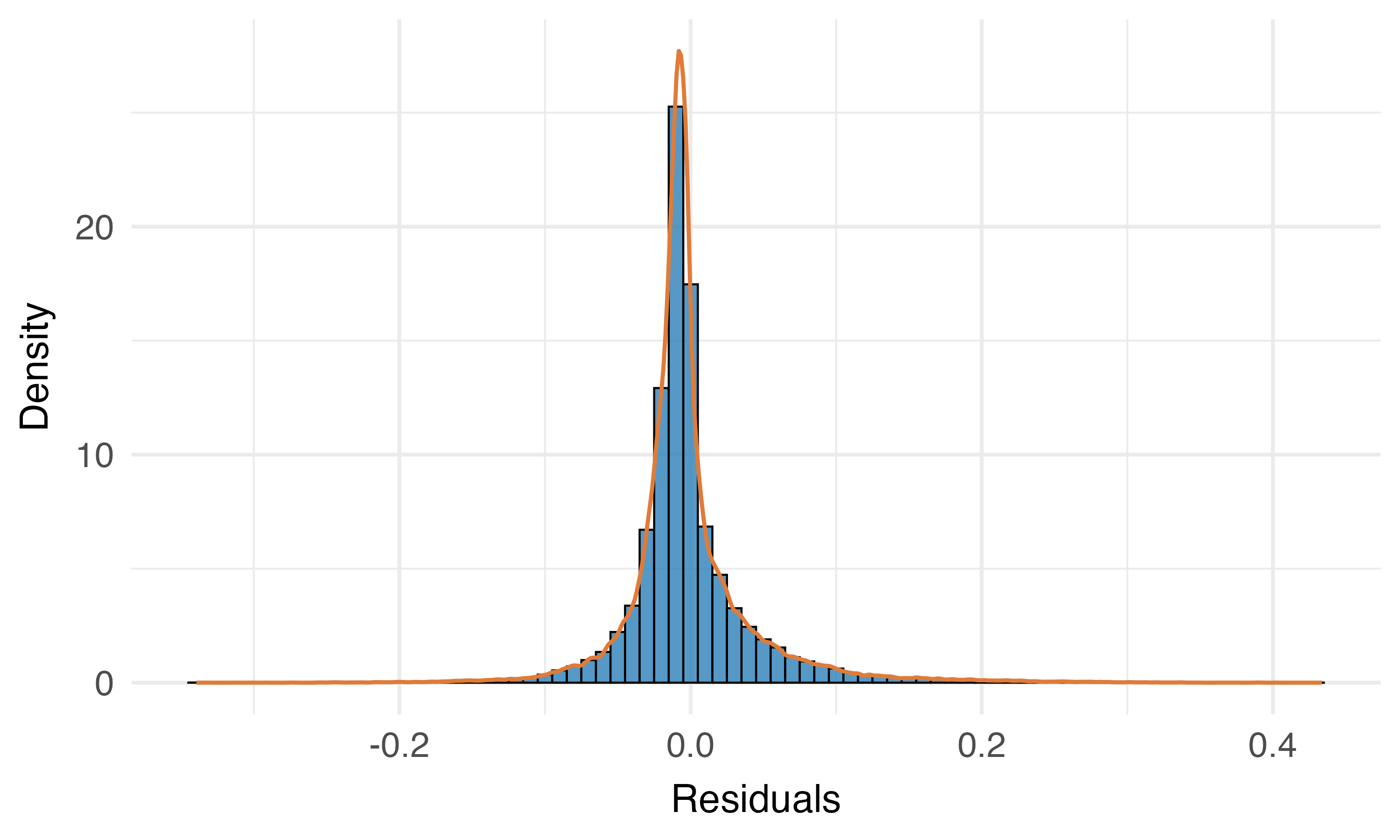}
        \caption{Histogram of residuals}
        \label{fig:histogram}
    \end{subfigure}
    \hfill
    \begin{subfigure}[b]{0.49\textwidth}
        \centering
        \includegraphics[width=\textwidth]{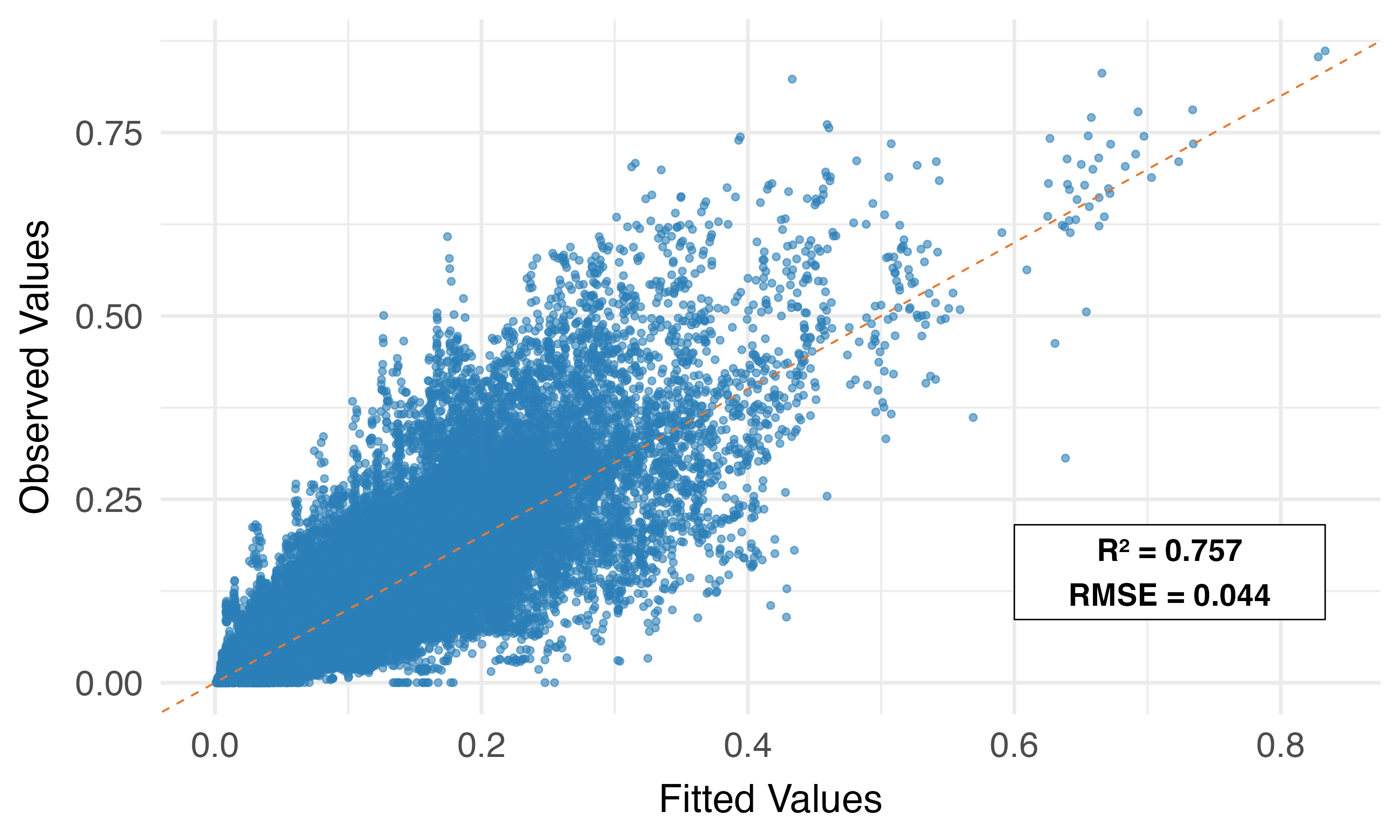}
        \caption{Fitted vs observed}
        \label{fig:fitted_vs_observed}
    \end{subfigure}
    
    \caption{Combined diagnostic plots for selected Beta regression GAMM model including smooth terms + random country level effects + interaction terms, and $R^2$ and RMSE on test performance metrics.}
    \label{fig:combined_diagnostics}
\end{figure}

The selected model's fit is also shown visually below in the histogram of residuals \ref{fig:histogram}, and observed versus fitted \ref{fig:fitted_vs_observed}. The RMSE (in units of workforce proportions, EPWA) on the test set is 0.044. See the Supplementary Figure \ref{fig:smooth_terms} for  detail on the non-linear relationships between predictors and the response, adjusted for country-specific random effects.

\subsubsection{Validation Tests}
We assessed performance across multiple validation types -- spatial, temporal, and multiscale (see Table \ref{tab:gam-validation-results}). The RMSE for spatial validation is 0.046, with that of the forward- and backward- predicting 0.045 and 0.054. We note that the same countries are used in these training examples, but they represent consistent ability for the model to spatially interpolate and forecast and hindcast within countries where data may be missing.

\begin{table}[ht]
\centering
\small
\caption{\label{tab:gam-validation-results}Root Mean Squared Error (RMSE) results for the GAM model across validation types and regions. Regions include: South America (SA), Southern Europe (SE), Eastern Europe (EE), Western Europe (WE), Central America (CA), Northern Europe (NE), East Asia (EA), Northern America (NA), and Australia and New Zealand (ANZ).}
\centering
\begin{tabular}[t]{lr}
\toprule
\textbf{Validation} & \textbf{RMSE}\\
\midrule
\cellcolor{gray!10}{Time Backward} & \cellcolor{gray!10}{0.054}\\
Time Forward & 0.045\\
\cellcolor{gray!10}{Spatial} & \cellcolor{gray!10}{0.046}\\
Multiscale: SA & 0.103\\
\cellcolor{gray!10}{Multiscale: SE} & \cellcolor{gray!10}{0.091}\\
Multiscale: EE & 0.092\\
\cellcolor{gray!10}{Multiscale: WE} & \cellcolor{gray!10}{0.035}\\
Multiscale: CA & 0.089\\
\cellcolor{gray!10}{Multiscale: NE} & \cellcolor{gray!10}{0.036}\\
Multiscale: EA & 0.048\\
\cellcolor{gray!10}{Multiscale: NA} & \cellcolor{gray!10}{0.079}\\
Multiscale: ANZ & 0.033\\
\bottomrule
\end{tabular}
\end{table}

\normalsize
The multiscale validation results show an increase in RMSE, as may be expected, with values typically around double those of the spatial and temporal validations. For instance, in Southern Europe (SE), the RMSE value rises to 0.0906. In general, the multiscale RMSE remaining is at most roughly double that of spatial and temporal validation on the same scale. Given the absence of ground-truth data for many subnational units, the multiscale validation approach is particularly valuable for understanding the error that may be expected for deploying at higher resolution. 

\subsection{Distribution of the agricultural workforce}

\begin{figure}[H]
    \centering
    \begin{subfigure}[ht]{\textwidth}
        \centering
        \includegraphics[width=0.7\textwidth]{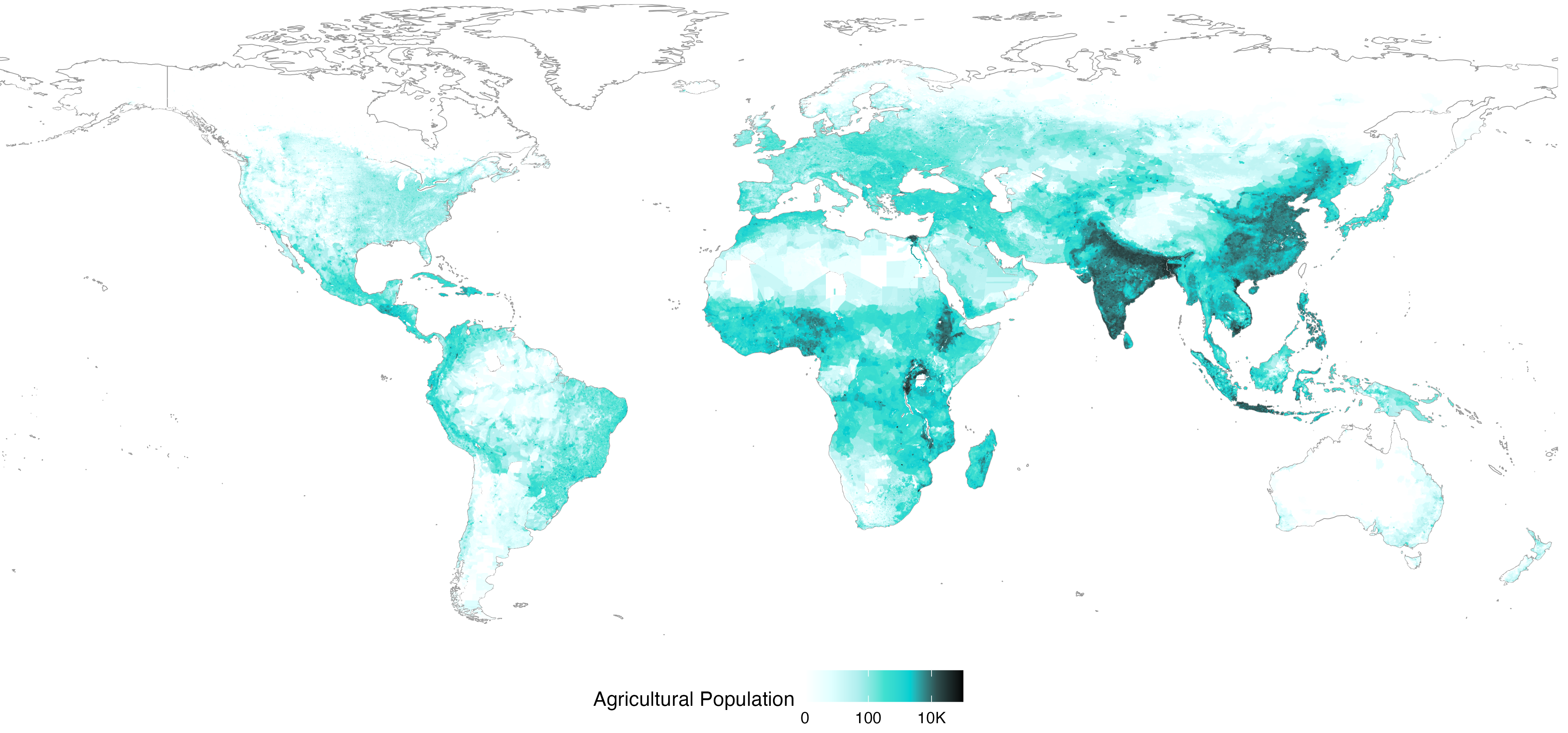}
        \caption{Predicted agricultural workforce year 2020.}
        \label{fig:subfig_pop2020}
    \end{subfigure}
    \vspace{0.5cm} 
    \begin{subfigure}[ht]{\textwidth}
        \centering
        \includegraphics[width=0.7\textwidth]{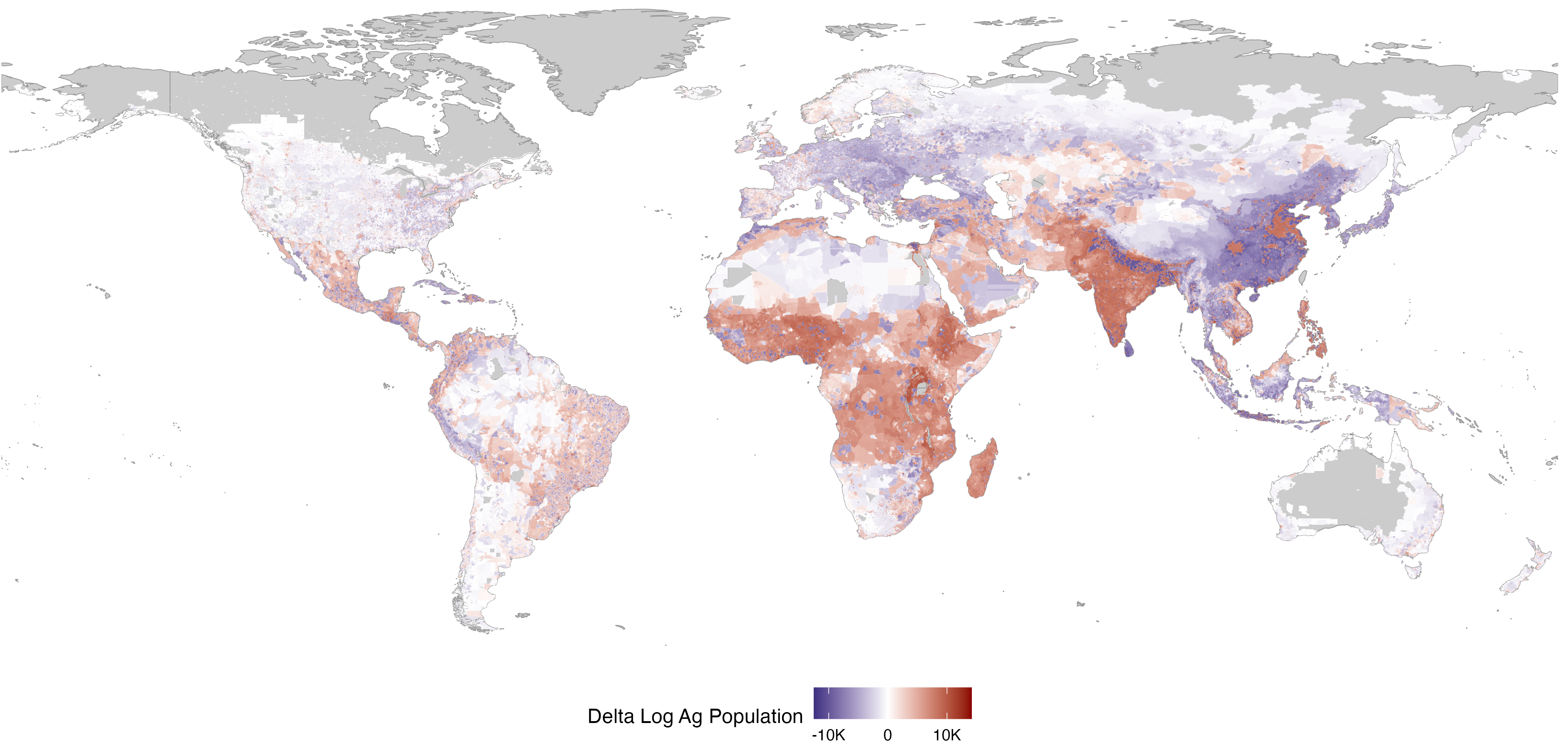}
        \caption{Projected change in agricultural workforce by 2050.}
        \label{fig:subfig_change2050}
    \end{subfigure}
    \vspace{0.1cm} 
    \begin{subfigure}[ht]{\textwidth}
        \centering
        \includegraphics[width=0.7\textwidth]{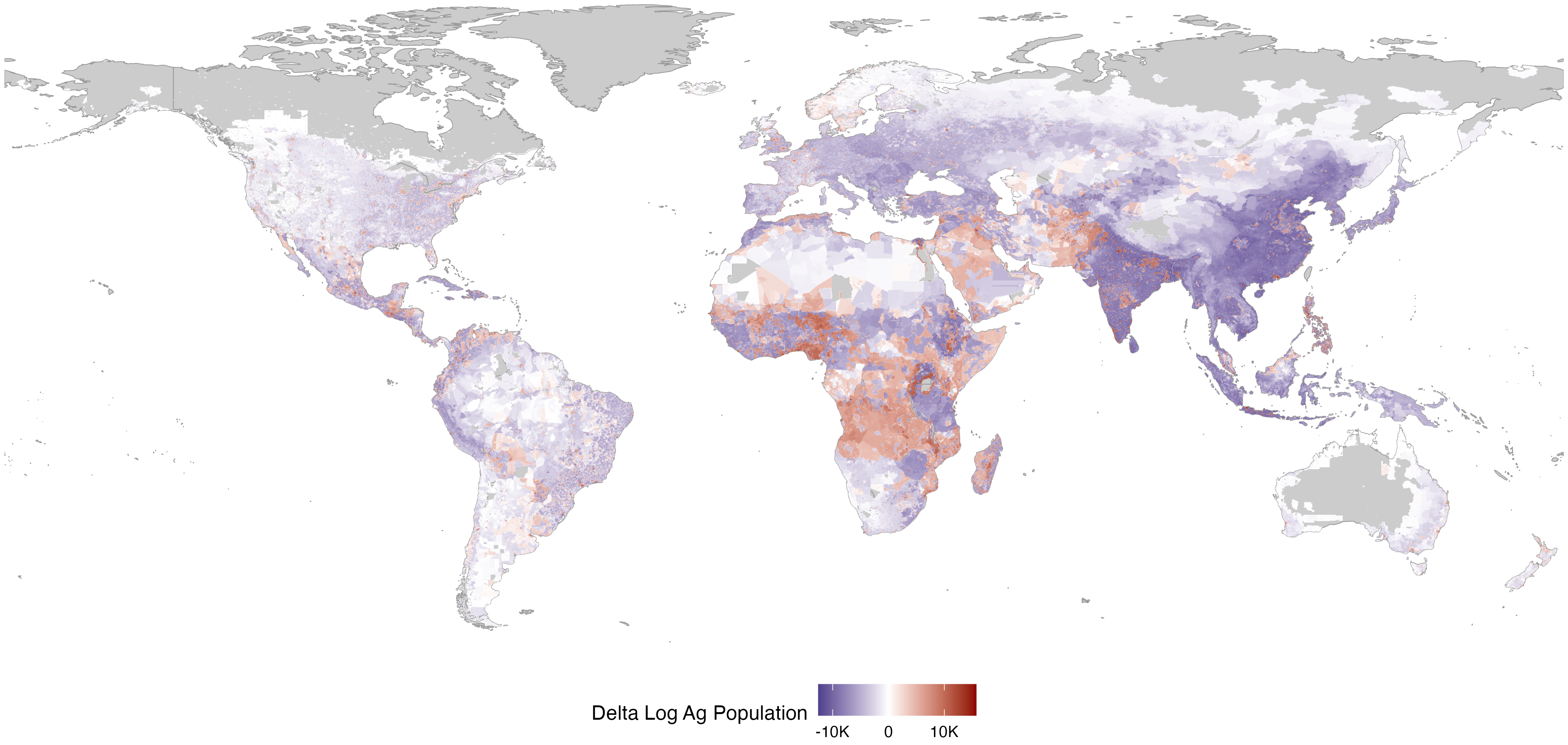}
        \caption{Projected change in agricultural workforce by 2100.}
        \label{fig:subfig_change2100}
    \end{subfigure}
    \caption{Current and future distribution of the agricultural workforce. (a) the estimated agricultural workforce distribution (log10 scale) in 2020. Figures (b) and (c) show the projected change in the agricultural workforce by 2050 and 2100, respectively, under SSP2. Positive values indicate an increase in agricultural workforce, while negative values represent a decline. The color gradient ranges from deep blue (significant reductions) to dark red (significant increases), with neutral changes represented in white. Regions in gray indicate missing data.}
    \label{fig:ag_pop_and_change}
\end{figure}

Maps for the agricultural workforce are shown in Figure \ref{fig:ag_pop_and_change} revealing both the existing spatial heterogeneity in this workforce in contemporary times 2020, as well as the spatial dynamics expected under socio-economic development in future by 2050 and 2100. Figures \ref{fig:subfig_gam2000}, \ref{fig:subfig_gam2050}, and \ref{fig:subfig_gam2100} show the downscaled ratios of employed population working in agriculture at a $0.083$ $\times0.083$ degrees resolution, for years 2000, 2050 and 2100, respectively (see Table \ref{tab:distribution_ag_workforce_2}, Tables \ref{tab:ag_pop_ssp1_uncorrected} to \ref{tab:ag_pop_ssp5_corrected} in the appendix for SSPs 1 to 5). In the baseline period (year 2020),  and under SSP2, we find globally 1.22 billion were employed in agriculture (96\% of the global population represented\footnote{The land/sea mask mismatch across agricultural lands and population layers on the coastline, resulting in omitted coastal populations -- mainly affects urban, densely populated areas. We also convert to 0 any grid cell where the population is less than 1 but greater than 0 due to interpolation techniques in the population layers.}). Under the business as usual scenario (SSP2), this figure is expected to grow to 1.33 billion by 2050, and then fall to 1.06 billion by 2100. 

\begingroup\fontsize{8}{10}\selectfont

\begin{longtable}[t]{llllll}
\caption{\label{tab:distribution_ag_workforce_2}Uncorrected Predictions of Total Agricultural Population and Percent Change by World Bank Region under scenario SSP2}\\
\toprule
WB Region & Pop baseline & Pop SSP2 2050 & $\Delta$ Pop SSP2 2050 & Pop SSP2 2100 & $\Delta$ Pop SSP2 2100\\
\midrule
\endfirsthead
\caption[]{Uncorrected Predictions of Total Agricultural Population and Percent Change by World Bank Region under scenario SSP2 \textit{(continued)}}\\
\toprule
WB Region & Pop baseline & Pop SSP2 2050 & $\Delta$ Pop SSP2 2050 & Pop SSP2 2100 & $\Delta$ Pop SSP2 2100\\
\midrule
\endhead

\endfoot
\bottomrule
\endlastfoot
\cellcolor{gray!10}{EAP} & \cellcolor{gray!10}{426.8M} & \cellcolor{gray!10}{381.5M} & \cellcolor{gray!10}{-10.6\%} & \cellcolor{gray!10}{254.1M} & \cellcolor{gray!10}{-40.5\%}\\
ECA & 36.5M & 32.1M & -12.1\% & 25.2M & -30.9\%\\
\cellcolor{gray!10}{LAC} & \cellcolor{gray!10}{43.0M} & \cellcolor{gray!10}{43.7M} & \cellcolor{gray!10}{1.6\%} & \cellcolor{gray!10}{41.0M} & \cellcolor{gray!10}{-4.5\%}\\
MENA & 37.3M & 38.8M & 4.0\% & 37.3M & 0.1\%\\
\cellcolor{gray!10}{NA} & \cellcolor{gray!10}{4.8M} & \cellcolor{gray!10}{4.5M} & \cellcolor{gray!10}{-4.8\%} & \cellcolor{gray!10}{4.0M} & \cellcolor{gray!10}{-16.3\%}\\
SA & 406.4M & 442.4M & 8.8\% & 349.9M & -13.9\%\\
\cellcolor{gray!10}{SSA} & \cellcolor{gray!10}{269.3M} & \cellcolor{gray!10}{387.8M} & \cellcolor{gray!10}{44.0\%} & \cellcolor{gray!10}{354.1M} & \cellcolor{gray!10}{31.5\%}\\*
\end{longtable}
\endgroup{}

There are notable regional differences. Again under SSP2, we find increases in agricultural workforce in South Asia (SA), Middle-East and North Africa (MENA) and Sub-Saharan Africa (SSA), particularly over 2020-2050. The highest increase is projected for SSA (44.0\% increase) by 2050. Conversely, we find decreases in East Asia \& Pacific (-10.6\%) Europe \& Central Asia (ECA), and North America (NA). There is, however, significant subregional and subnational heterogeneity. In contrast, Latin America and the Caribbean (LAC) and SA exhibit initial increases in agricultural workforce by 2050 (1.6\% for LAC and 8.8\% for SA), but projections show a decline by 2100 (-4.5\% for LAC and -13.9\% for SA), showing a shift in trends over the long term.

\bigskip

When comparing these projections to the findings by Mehrabi (2023) \cite{mehrabi2023likely}, which projects the number and size of farms, we see that the magnitudes and signs of change are consistent. It is reasonable to expect that changes in the number of farms and the agricultural workforce should, more or less, go hand in hand. We have attempted to model the same structural factors, such as population growth,  economic development and rural-urban migration in   driving agricultural development, albeit through the lens of agricultural workers themselves, and at a very different scale, hence the correspondence between the results when aggregated to the regional level is reassuring.

\bigskip

Our results novelly show important differences in country-level workforce dynamics. Table \ref{tab:top_ag_changes2100} provides the top 20 countries for largest absolute increases and decreases by 2100 under SSP2 (Table \ref{tab:top_ag_changes2050} for changes by 2050). Of the top 20 increases, 14 belong to SSA (15 in 2050). By 2100, Nigeria's agricultural workforce is expected to increase by 33.61 million, the largest absolute increase. This is followed by Uganda and the Democratic Republic of the Congo with increases of 12.72 million and 8.85 million, respectively. These top three countries, along with other nations in SSA like Niger, Malawi, and Tanzania, underscore a consistent trend with the overall increase in agricultural reliance within the region, driven by population growth and continued dependence on agriculture as a primary economic activity.

\bigskip

The analysis of agricultural workforce changes projected for 2100 also reveals substantial declines in several key agricultural regions, notably within EAP and SA. In EAP, China reports the most significant reduction, with a projected decrease of approximately 153.14 million in the agricultural workforce, underscoring major shifts towards urbanization and industrial growth that reduce reliance on agricultural employment. Other countries in the region, such as Indonesia, Thailand, Viet Nam, Myanmar or Japan, are also expected to see substantial declines in agricultural populations. In SA, India, often seen as an agricultural stronghold and which shows an increase by 2050, appears to reverse this trend in the long term with one of the major decreases by 2100, of about 44.29 million, reflecting significant economic transformations and shifts in agricultural practices or technology adoption. Bangladesh, Nepal, and Sri Lanka are also among the top countries with the largest decreases by 2100. In other regions like LAC, MENA and Eastern ECA  notable reductions are similarly observed. For instance, Brazil and Peru in the LAC region see decreases of 1.20 million and 1.13 million respectively. Turkey, Morocco and Egypt are expected to show large decreases by 2050 and Turkey continues to project significant declines by 2100. In ECA, countries like the Russian Federation and Poland exhibit decreases, suggesting changes in agricultural dynamics possibly driven by economic policies or demographic shifts. 

\bigskip

Notably, large percentage changes can occur in places with relatively small absolute numbers (e.g., Kuwait, Bahrain). The changes that we observe in this selection of countries and, in particular, around peri-urban areas that see an increase on supply and agricultural activity as they grow, are consistent with different processes. This is related to the interaction effect that has potentially higher uncertainty and yields spatial differences, but has overall no impact on the actual main results.

\bigskip

There is important subnational heterogeneity as well. For example, while South Asia's agricultural workforce is likely to increase for the first part, several subnational units in Bangladesh and India are expected to see the greatest growths as well as losses. This can be linked to both internal migration and fluctuating agricultural intensities, as certain regions become more urbanized and adopt more labour-efficient farming methods, while others experience stronger increases in rural population growth. In contrast, some countries with high expected rates of urbanization and economic growth—China, Eastern European countries—show more consistent declines across most subnational units. The data also illustrates the broader regional impacts, with countries from EAP like Indonesia, Philippines, and Myanmar appearing in both the positive and negative change categories. This dual presence highlights the diverse economic transitions within the region, affecting agricultural labour differently across countries.

\section{Conclusions}  
The agricultural workforce is a critical component of global food security, economic stability, and rural development. Understanding the current state, dynamics, and future trends of this workforce is essential, particularly to assess the risks that they face, including but not limiting to exposure to pollution and climate hazards, diseases, extreme weather and other socioeconomic shocks. In this study, we present the first global assessment and data set detailing current and future agricultural workforce for public use. The selected GAMM model includes smooth terms and interaction effects, capturing major trends but also significantly improving performance by capturing non-linear relationships and regional variations. We expect this dataset to have a wide range of downstream uses and applications and release the full code and data with this paper.

\bigskip

Our main results highlight several key trends. Firstly, we map dynamics in the proportion of the population engaged in agriculture, with declines particularly in upper-middle income countries and some lower-middle income countries, driven by technological advancements and shifts towards more urbanized economies that are currently happening and will continue to happen in the near future. High-income countries see more nuanced decreases of the agricultural population, representing a stabilization of the sector. Conversely, we show that certain regions, particularly in low-income countries, will likely experience increases in demand for agricultural employment in the future, due to slower economic transitions and ongoing reliance on agriculture for livelihoods. This increased need for job generation requires address and may have been overlooked to date in policy.

\bigskip

Downscaled data on the agricultural workforce is necessary for different applications. In epidemiological studies, such data can help pinpoint where agricultural populations are concentrated, aiding in the identification and mitigation of health risks such as pesticide exposure \cite{fuhrimann2019exposure, barron_cuenca_pesticide_2020}, infectious diseases or disease outbreaks \cite{graham2008animal, withers2002antibody}. In climate change analysis, high-resolution workforce data could be crucial for assessing the impact of heat stress \cite{nelson2024global}, flooding, droughts, and more, as agricultural workers are among the most vulnerable to the changing climate. Additionally, detailed data could support disaster preparedness and response, enabling more accurate socio-economic planning and policy-making, and enhances the effectiveness of development programs targeting rural and agricultural communities \cite{nowak_opportunities_2024}.

\bigskip

It is our hope that by providing an easily accessible granular view of the current and future agricultural workforce, we can better understand and address the challenges faced by this critical segment of society who are so fundamental to feeding everyone on the planet.

\bigskip
\bigskip

\noindent \textbf{Code and data availability}: The data outputs from this study, including those of agricultural workforce projections, both corrected and uncorrected at the available units, are available in Geotiff format at link, alongside usage notes [publicly available on publication at DOI: \href{https://doi.org/10.5281/zenodo.14443333}{10.5281/zenodo.14443333}]. Scripts and model code developed for the analysis, including those used to train and project agricultural workforce changes, are also stored on the Zenodo repository and forks may be made from the corresponding author's Github repository [add on proofing].

\bigskip

\noindent \textbf{Author contribution:} NOZ and ZM designed and planned the study with SM providing input on model selection and validation. 
NOZ collected and harmonized the data, wrote the modeling pipelines and conducted the downstream analysis. 
NOZ , ZM and SM interpreted the results.
NOZ wrote the first draft of the paper with input from ZM. 
NOZ, ZM and SM revised the paper.

\bigskip

\noindent \textbf{Competing interests:} The authors declare no conflict of interest. 

\bigskip

\noindent \textbf{Acknowledgments:} NOZ was funded from a fellowship from ``la Caixa'' Foundation (ID 100010434), fellowship code LCF/BQ/EU22/11930082 for the period August 2023 - July 2025.  NOZ would also like to extend her gratitude to the LUGE Lab and Professor Navin Ramankutty for the opportunity to present the work and for their invaluable insights, which have significantly enriched this work.


\newpage
\bibliography{mybib}

\newpage
\appendix
\section*{Appendix}
\renewcommand{\thefigure}{\thesection\arabic{figure}}
\setcounter{figure}{0}
\section{Label sources}
\label{sec:labels}
Subnational statistics are given in Table \ref{tab:subnational_labels}, noting all the countries, years availability, and sources of data. The national data was retrieved from \href{https://ilostat.ilo.org/data/data-explorer/}{ILO Data Explorer} and Table \ref{tab:national_labels_1} shows the available years for these. However, data for China, which was retrieved from the Ministry of Agriculture and Rural Affairs of the People's Republic of China, provides data on agricultural employment vs. total employment (\%) from 'China Agricultural Development Report. China Statistical Communiqué of the National Economic and Social Development 2015\footnote{This source was used for China as the ILO Data Explorer only has one data point for China, corresponding to the year 2000, which aligns with figures from 1980-1985 according to the ministry's report. The report can be accessed at \url{http://english.moa.gov.cn/overview/201910/t20191009_296610.html}}.

\begin{table}[H]
\centering
\caption{Summary of availability of subnational data across years and sources.} 
\scriptsize
\begin{tabular}{lllrll}
  \hline
 Country & Start Year & End Year & Years & Missing Years & Source and Link \\ 
   \hline
 Australia & 2011 & 2016 &   2 & TRUE  & \href{https://explore.data.abs.gov.au/}{Australian Bureau of Statistics} \\ 
 Austria & 2000 & 2020 &  21 & FALSE & \href{https://ec.europa.eu/eurostat/databrowser/view/lfst_r_lfe2en2/default/table?lang=en}{Eurostat} \\ 
 Belgium & 2003 & 2020 &  18 & FALSE & \href{https://ec.europa.eu/eurostat/databrowser/view/lfst_r_lfe2en2/default/table?lang=en}{Eurostat} \\ 
 Brazil & 2001 & 2015 &  14 & TRUE  & \href{https://www.ibge.gov.br/en/statistics/multi-domain/science-technology-and-innovation/20620-summary-of-indicators-pnad2.html?=&t=microdados}{IBGE} \\ 
 Bulgaria & 2000 & 2020 &  21 & FALSE & \href{https://ec.europa.eu/eurostat/databrowser/view/lfst_r_lfe2en2/default/table?lang=en}{Eurostat} \\ 
 Canada & 2006 & 2020 &  15 & FALSE & \href{https://www150.statcan.gc.ca/t1/tbl1/en/cv.action?pid=1410037901}{Statistics Canada} \\ 
 Chile & 2010 & 2019 &  10 & FALSE & \href{https://bancodatosene.ine.cl/}{INE Chile} \\ 
 Colombia & 2015 & 2020 &   6 & FALSE & \href{https://www.dane.gov.co/index.php/estadisticas-por-tema/mercado-laboral/mercado-laboral-por-departamentos/mercado-laboral-por-departamento-historicos}{DANE Colombia} \\ 
 Croatia & 2000 & 2020 &  21 & FALSE & \href{https://ec.europa.eu/eurostat/databrowser/view/lfst_r_lfe2en2/default/table?lang=en}{Eurostat} \\ 
 Cyprus & 2000 & 2020 &  21 & FALSE & \href{https://ec.europa.eu/eurostat/databrowser/view/lfst_r_lfe2en2/default/table?lang=en}{Eurostat} \\ 
 Czechia & 2000 & 2020 &  21 & FALSE & \href{https://ec.europa.eu/eurostat/databrowser/view/lfst_r_lfe2en2/default/table?lang=en}{Eurostat} \\ 
 Denmark & 2000 & 2020 &  21 & FALSE & \href{https://ec.europa.eu/eurostat/databrowser/view/lfst_r_lfe2en2/default/table?lang=en}{Eurostat} \\ 
 Estonia & 2000 & 2020 &  21 & FALSE & \href{https://ec.europa.eu/eurostat/databrowser/view/lfst_r_lfe2en2/default/table?lang=en}{Eurostat} \\ 
 Finland & 2000 & 2020 &  21 & FALSE & \href{https://ec.europa.eu/eurostat/databrowser/view/lfst_r_lfe2en2/default/table?lang=en}{Eurostat} \\ 
 France & 2000 & 2020 &  21 & FALSE & \href{https://ec.europa.eu/eurostat/databrowser/view/lfst_r_lfe2en2/default/table?lang=en}{Eurostat} \\ 
 Germany & 2000 & 2020 &  21 & FALSE & \href{https://ec.europa.eu/eurostat/databrowser/view/lfst_r_lfe2en2/default/table?lang=en}{Eurostat} \\ 
 Greece & 2000 & 2020 &  21 & FALSE & \href{https://ec.europa.eu/eurostat/databrowser/view/lfst_r_lfe2en2/default/table?lang=en}{Eurostat} \\ 
 Hungary & 2000 & 2020 &  21 & FALSE & \href{https://ec.europa.eu/eurostat/databrowser/view/lfst_r_lfe2en2/default/table?lang=en}{Eurostat} \\ 
 Ireland & 2000 & 2020 &  21 & FALSE & \href{https://ec.europa.eu/eurostat/databrowser/view/lfst_r_lfe2en2/default/table?lang=en}{Eurostat} \\ 
 Italy & 2000 & 2020 &  21 & FALSE & \href{https://ec.europa.eu/eurostat/databrowser/view/lfst_r_lfe2en2/default/table?lang=en}{Eurostat} \\ 
 Latvia & 2000 & 2020 &  21 & FALSE & \href{https://ec.europa.eu/eurostat/databrowser/view/lfst_r_lfe2en2/default/table?lang=en}{Eurostat} \\ 
 Lithuania & 2000 & 2020 &  21 & FALSE & \href{https://ec.europa.eu/eurostat/databrowser/view/lfst_r_lfe2en2/default/table?lang=en}{Eurostat} \\ 
 Luxembourg & 2000 & 2020 &  21 & FALSE & \href{https://ec.europa.eu/eurostat/databrowser/view/lfst_r_lfe2en2/default/table?lang=en}{Eurostat} \\ 
 Malta & 2000 & 2020 &  21 & FALSE & \href{https://ec.europa.eu/eurostat/databrowser/view/lfst_r_lfe2en2/default/table?lang=en}{Eurostat} \\ 
 Mexico & 2005 & 2020 &  16 & FALSE & \href{https://www.inegi.org.mx/programas/enoe/15ymas/#datos_abiertos}{INEGI Mexico} \\ 
 Netherlands & 2000 & 2020 &  21 & FALSE & \href{https://ec.europa.eu/eurostat/databrowser/view/lfst_r_lfe2en2/default/table?lang=en}{Eurostat} \\ 
 North Macedonia & 2000 & 2020 &  21 & FALSE & \href{https://ec.europa.eu/eurostat/databrowser/view/lfst_r_lfe2en2/default/table?lang=en}{Eurostat} \\ 
 Norway & 2008 & 2020 &  13 & FALSE & \href{https://ec.europa.eu/eurostat/databrowser/view/lfst_r_lfe2en2/default/table?lang=en}{Eurostat} \\ 
 Poland & 2000 & 2020 &  21 & FALSE & \href{https://ec.europa.eu/eurostat/databrowser/view/lfst_r_lfe2en2/default/table?lang=en}{Eurostat} \\ 
 Portugal & 2000 & 2020 &  21 & FALSE & \href{https://ec.europa.eu/eurostat/databrowser/view/lfst_r_lfe2en2/default/table?lang=en}{Eurostat} \\ 
 Romania & 2000 & 2020 &  21 & FALSE & \href{https://ec.europa.eu/eurostat/databrowser/view/lfst_r_lfe2en2/default/table?lang=en}{Eurostat} \\ 
 Serbia & 2000 & 2020 &  21 & FALSE & \href{https://ec.europa.eu/eurostat/databrowser/view/lfst_r_lfe2en2/default/table?lang=en}{Eurostat} \\ 
 Slovakia & 2000 & 2020 &  21 & FALSE & \href{https://ec.europa.eu/eurostat/databrowser/view/lfst_r_lfe2en2/default/table?lang=en}{Eurostat} \\ 
 Slovenia & 2000 & 2020 &  21 & FALSE & \href{https://ec.europa.eu/eurostat/databrowser/view/lfst_r_lfe2en2/default/table?lang=en}{Eurostat} \\ 
 Spain & 2000 & 2020 &  21 & FALSE & \href{https://ec.europa.eu/eurostat/databrowser/view/lfst_r_lfe2en2/default/table?lang=en}{Eurostat} \\ 
 Sweden & 2000 & 2020 &  21 & FALSE & \href{https://ec.europa.eu/eurostat/databrowser/view/lfst_r_lfe2en2/default/table?lang=en}{Eurostat} \\ 
 United States & 2001 & 2020 &  20 & FALSE & \href{https://www.bea.gov/data/employment}{BEA USA} \\ 
  \hline
  \end{tabular}
\label{tab:subnational_labels}
\end{table}

\begingroup\fontsize{8}{10}\selectfont

\begin{longtable}[t]{lcccc}
\caption{\label{tab:national_labels_1}Summary of availability of subnational data across years}\\
\toprule
country & start\_year & end\_year & years & missing\_years\\
\midrule
\endfirsthead
\caption[]{Summary of availability of subnational data across years \textit{(continued)}}\\
\toprule
country & start\_year & end\_year & years & missing\_years\\
\midrule
\endhead

\endfoot
\bottomrule
\endlastfoot
Afghanistan & 2008 & 2020 & 5 & TRUE\\
Albania & 2007 & 2019 & 13 & FALSE\\
Angola & 2004 & 2019 & 5 & TRUE\\
Argentina & 2004 & 2020 & 14 & TRUE\\
Armenia & 2007 & 2020 & 14 & FALSE\\
\addlinespace
Bangladesh & 2006 & 2017 & 4 & TRUE\\
Barbados & 2015 & 2019 & 5 & FALSE\\
Belarus & 2016 & 2020 & 5 & FALSE\\
Belize & 2014 & 2019 & 5 & TRUE\\
Benin & 2011 & 2019 & 2 & TRUE\\
\addlinespace
Bhutan & 2018 & 2020 & 3 & FALSE\\
Bolivia (Plurinational State of) & 2000 & 2020 & 18 & TRUE\\
Bosnia and Herzegovina & 2001 & 2020 & 16 & TRUE\\
Botswana & 2006 & 2020 & 4 & TRUE\\
Brunei Darussalam & 2014 & 2020 & 5 & TRUE\\
\addlinespace
Burkina Faso & 2014 & 2018 & 2 & TRUE\\
Burundi & 2014 & 2020 & 2 & TRUE\\
Cabo Verde & 2009 & 2015 & 2 & TRUE\\
Cambodia & 2000 & 2020 & 16 & TRUE\\
Cameroon & 2007 & 2014 & 2 & TRUE\\
\addlinespace
Chad & 2018 & 2018 & 1 & FALSE\\
China & 2000 & 2000 & 1 & FALSE\\
Comoros & 2004 & 2014 & 2 & TRUE\\
Congo & 2005 & 2009 & 2 & TRUE\\
Congo, Democratic Republic of the & 2005 & 2012 & 2 & TRUE\\
\addlinespace
Cook Islands & 2016 & 2019 & 2 & TRUE\\
Costa Rica & 2001 & 2020 & 20 & FALSE\\
Côte d?Ivoire & 2012 & 2019 & 5 & TRUE\\
Djibouti & 2017 & 2017 & 1 & FALSE\\
Dominican Republic & 2000 & 2020 & 21 & FALSE\\
\addlinespace
Ecuador & 2001 & 2020 & 19 & TRUE\\
Egypt & 2008 & 2020 & 12 & TRUE\\
El Salvador & 2010 & 2020 & 11 & FALSE\\
Eswatini & 2016 & 2016 & 1 & FALSE\\
Ethiopia & 2005 & 2013 & 2 & TRUE\\
\addlinespace
Fiji & 2005 & 2016 & 3 & TRUE\\
Gambia & 2012 & 2018 & 2 & TRUE\\
Georgia & 2009 & 2020 & 12 & FALSE\\
Ghana & 2000 & 2017 & 6 & TRUE\\
Grenada & 2018 & 2020 & 3 & FALSE\\
\addlinespace
Guatemala & 2003 & 2019 & 13 & TRUE\\
Guinea & 2002 & 2002 & 1 & FALSE\\
Guinea-Bissau & 2018 & 2019 & 2 & FALSE\\
Guyana & 2018 & 2019 & 2 & FALSE\\
Haiti & 2012 & 2012 & 1 & FALSE\\
\addlinespace
Honduras & 2005 & 2020 & 16 & FALSE\\
Iceland & 2000 & 2020 & 21 & FALSE\\
India & 2000 & 2020 & 7 & TRUE\\
Indonesia & 2000 & 2020 & 21 & FALSE\\
Iran (Islamic Republic of) & 2005 & 2020 & 16 & FALSE\\
\addlinespace
Israel & 2012 & 2020 & 9 & FALSE\\
Jamaica & 2008 & 2020 & 12 & TRUE\\
Japan & 2000 & 2020 & 21 & FALSE\\
Jordan & 2017 & 2020 & 4 & FALSE\\
Kenya & 2019 & 2019 & 1 & FALSE\\
\addlinespace
Kiribati & 2015 & 2020 & 3 & TRUE\\
Kosovo & 2000 & 2020 & 10 & TRUE\\
Kyrgyzstan & 2010 & 2020 & 11 & FALSE\\
Lao People's Democratic Republic & 2010 & 2017 & 2 & TRUE\\
Lebanon & 2019 & 2019 & 1 & FALSE\\
\addlinespace
Lesotho & 2019 & 2019 & 1 & FALSE\\
Liberia & 2010 & 2017 & 2 & TRUE\\
Madagascar & 2001 & 2015 & 3 & TRUE\\
Malawi & 2013 & 2013 & 1 & FALSE\\
Malaysia & 2000 & 2000 & 1 & FALSE\\
\addlinespace
Maldives & 2009 & 2019 & 4 & TRUE\\
Mali & 2013 & 2020 & 7 & TRUE\\
Marshall Islands & 2019 & 2019 & 1 & FALSE\\
Mauritania & 2017 & 2019 & 2 & TRUE\\
Mauritius & 2001 & 2020 & 19 & TRUE\\
\addlinespace
Micronesia (Federated States of) & 2014 & 2014 & 1 & FALSE\\
Mongolia & 2003 & 2020 & 15 & TRUE\\
Montenegro & 2011 & 2020 & 10 & FALSE\\
Montserrat & 2020 & 2020 & 1 & FALSE\\
Morocco & 2004 & 2014 & 2 & TRUE\\
\addlinespace
Mozambique & 2015 & 2015 & 1 & FALSE\\
Myanmar & 2015 & 2020 & 5 & TRUE\\
Namibia & 2010 & 2018 & 6 & TRUE\\
Nauru & 2013 & 2013 & 1 & FALSE\\
Nepal & 2008 & 2017 & 2 & TRUE\\
\addlinespace
New Caledonia & 2017 & 2020 & 4 & FALSE\\
Nicaragua & 2001 & 2014 & 3 & TRUE\\
Niger & 2011 & 2019 & 4 & TRUE\\
Nigeria & 2011 & 2019 & 3 & TRUE\\
Niue & 2015 & 2017 & 2 & TRUE\\
\addlinespace
Occupied Palestinian Territory & 2000 & 2020 & 21 & FALSE\\
Pakistan & 2005 & 2019 & 14 & TRUE\\
Palau & 2014 & 2020 & 2 & TRUE\\
Panama & 2003 & 2020 & 15 & TRUE\\
Papua New Guinea & 2010 & 2010 & 1 & FALSE\\
\addlinespace
Paraguay & 2001 & 2020 & 20 & FALSE\\
Peru & 2000 & 2020 & 21 & FALSE\\
Philippines & 2003 & 2020 & 10 & TRUE\\
Republic of Korea & 2000 & 2020 & 21 & FALSE\\
Republic of Moldova & 2000 & 2020 & 21 & FALSE\\
\addlinespace
Russian Federation & 2010 & 2020 & 11 & FALSE\\
Rwanda & 2014 & 2020 & 5 & TRUE\\
Saint Lucia & 2017 & 2020 & 4 & FALSE\\
Samoa & 2012 & 2017 & 2 & TRUE\\
Sao Tome and Principe & 2017 & 2017 & 1 & FALSE\\
\addlinespace
Senegal & 2011 & 2019 & 4 & TRUE\\
Seychelles & 2014 & 2020 & 7 & FALSE\\
Sierra Leone & 2014 & 2018 & 2 & TRUE\\
Solomon Islands & 2005 & 2013 & 2 & TRUE\\
Somalia & 2019 & 2019 & 1 & FALSE\\
\addlinespace
South Africa & 2000 & 2020 & 21 & FALSE\\
South Sudan & 2008 & 2008 & 1 & FALSE\\
Sri Lanka & 2010 & 2020 & 10 & TRUE\\
Sudan & 2008 & 2011 & 2 & TRUE\\
Suriname & 2016 & 2016 & 1 & FALSE\\
\addlinespace
Switzerland & 2000 & 2020 & 21 & FALSE\\
Tajikistan & 2003 & 2009 & 3 & TRUE\\
Tanzania, United Republic of & 2001 & 2020 & 7 & TRUE\\
Thailand & 2000 & 2020 & 11 & TRUE\\
Timor-Leste & 2001 & 2016 & 5 & TRUE\\
\addlinespace
Togo & 2006 & 2019 & 5 & TRUE\\
Tonga & 2018 & 2018 & 1 & FALSE\\
Trinidad and Tobago & 2011 & 2020 & 8 & TRUE\\
Tunisia & 2005 & 2019 & 14 & TRUE\\
Tuvalu & 2016 & 2017 & 2 & FALSE\\
\addlinespace
Türkiye & 2000 & 2020 & 21 & FALSE\\
Uganda & 2010 & 2019 & 6 & TRUE\\
Ukraine & 2018 & 2020 & 3 & FALSE\\
United Arab Emirates & 2017 & 2020 & 4 & FALSE\\
United Kingdom of Great Britain and Northern Ireland & 2000 & 2020 & 21 & FALSE\\
\addlinespace
Uruguay & 2000 & 2020 & 21 & FALSE\\
Uzbekistan & 2020 & 2020 & 1 & FALSE\\
Vanuatu & 2006 & 2020 & 5 & TRUE\\
Venezuela (Bolivarian Republic of) & 2005 & 2017 & 10 & TRUE\\
Viet Nam & 2007 & 2020 & 13 & TRUE\\
\addlinespace
Wallis and Futuna & 2018 & 2019 & 2 & FALSE\\
Yemen & 2014 & 2014 & 1 & FALSE\\
Zambia & 2015 & 2020 & 5 & TRUE\\
Zimbabwe & 2011 & 2019 & 3 & TRUE\\*
\end{longtable}
\endgroup{}

\section{Population interpolation} \label{popinterpolate}
To interpolate population data for rural and total populations from raster files, an exponential growth model was used to estimate values for each year from 2000 to 2020 \cite{CIESIN2016}. The raster files provided data for the years 2000, 2010, and 2020, and were loaded containing both rural and total population data at a $0.0083 \times 0.0083$ degrees resolution.

\bigskip

For each pair of known years (2000-2010 and 2010-2020), we first compute the annualized growth rate between the rasters for two consecutive known years. This is given by:

\begin{equation}
    r = \frac{\ln(P_{t2} / P_{t1})}{t2 - t1}
\end{equation}
where $P_{t1}$ and $P_{t2}$ are the populations in the earlier and later years, respectively, and $t1$ and $t2$ are the years.

\bigskip

For each target year, the interpolated population was calculated using the exponential growth model:

\begin{equation}
    P_x = P_{t1} \cdot e^{r \cdot (t - t1)}
\end{equation}
where $P_x$ is the estimated population in the target year $t$, $P_{t1}$ is the population at the earlier known year, and $r$ is the annualized growth rate. This method assumes population changes follow an exponential trend rather than a linear one, which can more accurately reflect demographic dynamics over time.

\section{Additional experiments}
\label{sec:additional_exp}
We conducted a number of experiments to test the perfomance of alternative models, namely quantile generalized additive models (qGAM) and tree based methods, namely boosted regression trees and random forests in our pipeline. The results of these experiments are given below, alongside reasoning for why we did not employ these models in the final pipeline.

\subsection{qGAMs}

\begin{figure}[H]
    \centering
    \includegraphics[width=0.9\textwidth]{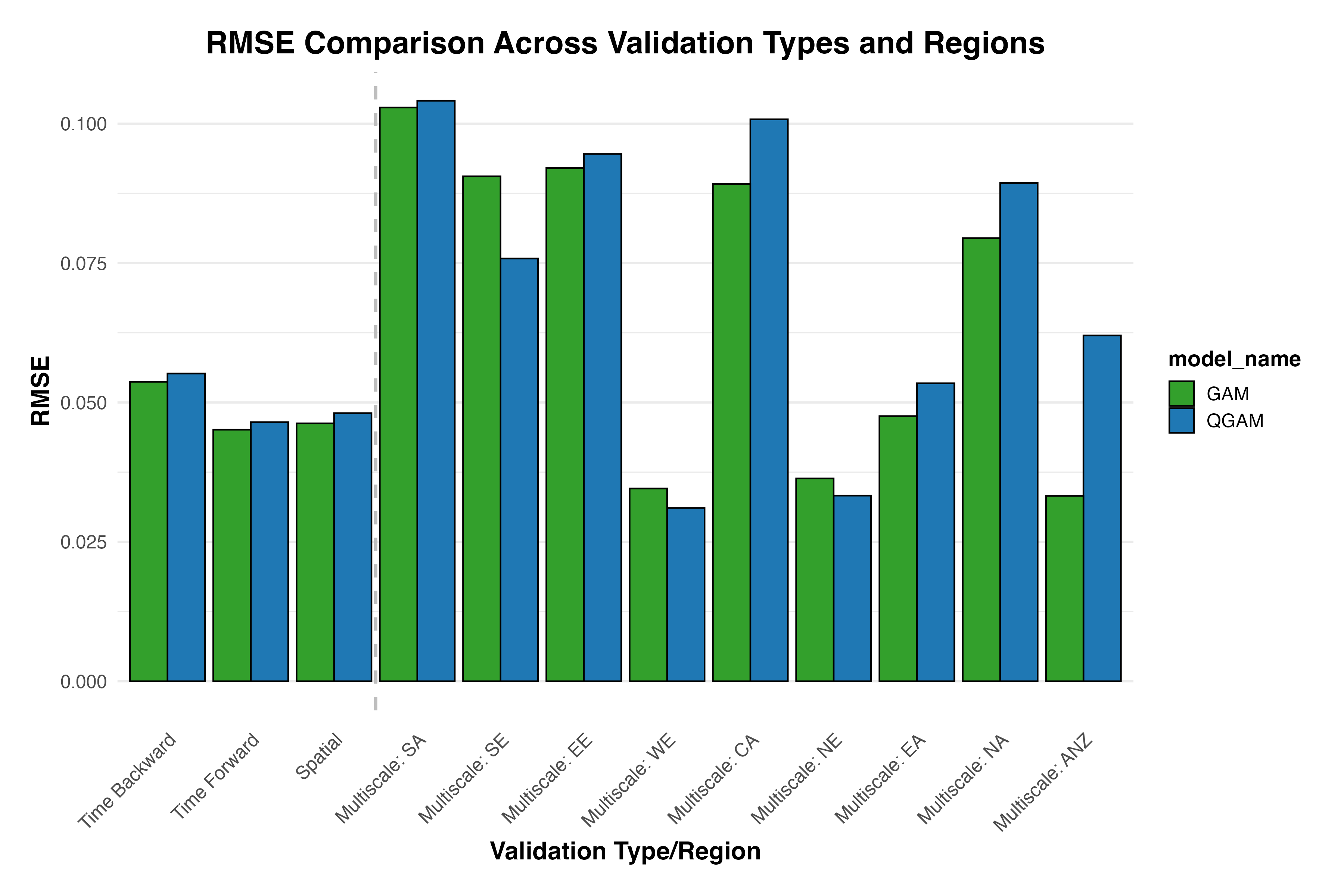}
    \caption{Validation strategies test results for the interactions qGAM and GAMM.}
    \label{fig:rmse_validation}
\end{figure}

\begin{figure}[H]
    \centering
    \begin{subfigure}[t]{\textwidth}
        \centering
        \includegraphics[width=0.9\textwidth]{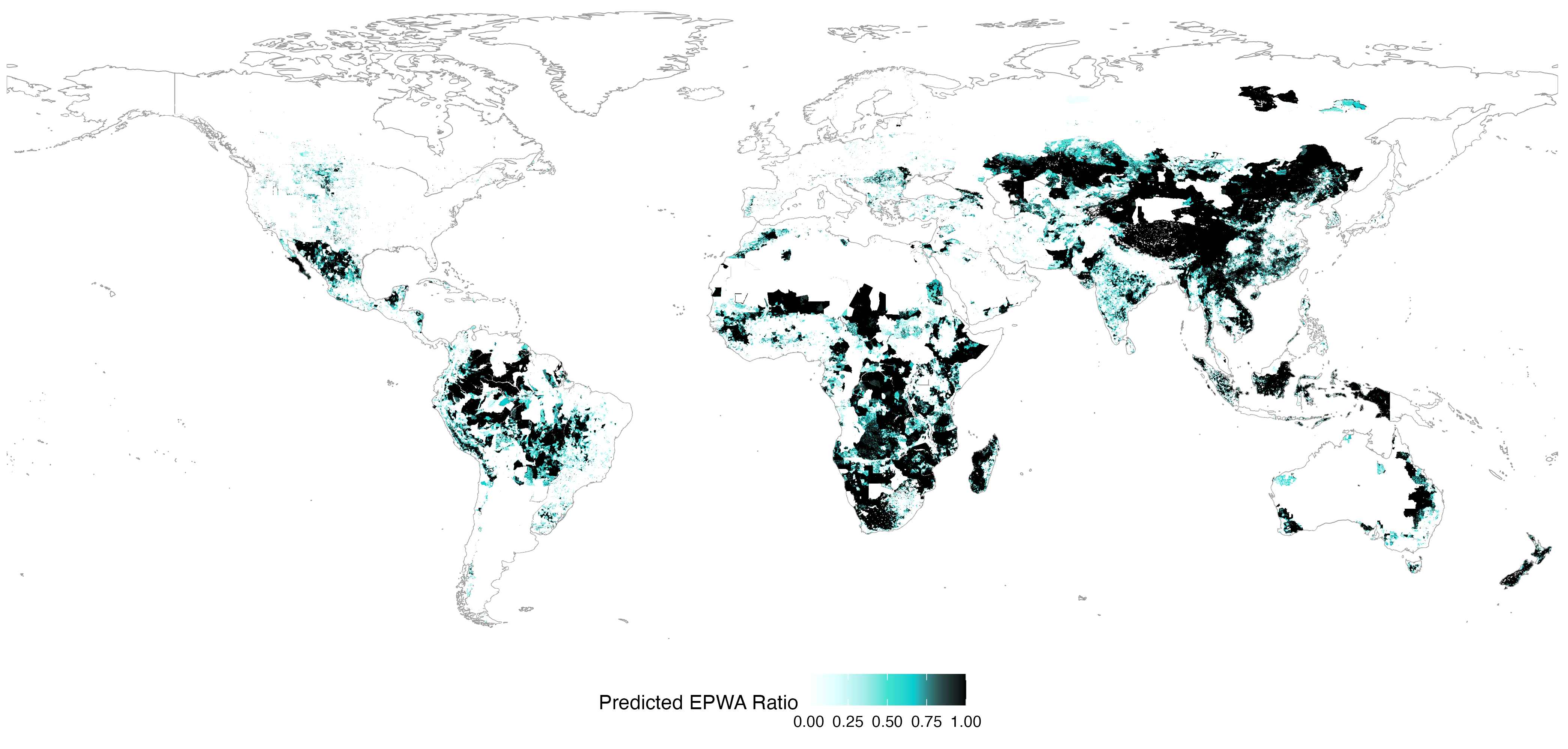}
        \caption{qGAM predictions for 2050.}
        \label{fig:SI_subfig_qgam}
    \end{subfigure}
    \vspace{0.5cm} 
    \begin{subfigure}[t]{\textwidth}
        \centering
        \includegraphics[width=0.9\textwidth]{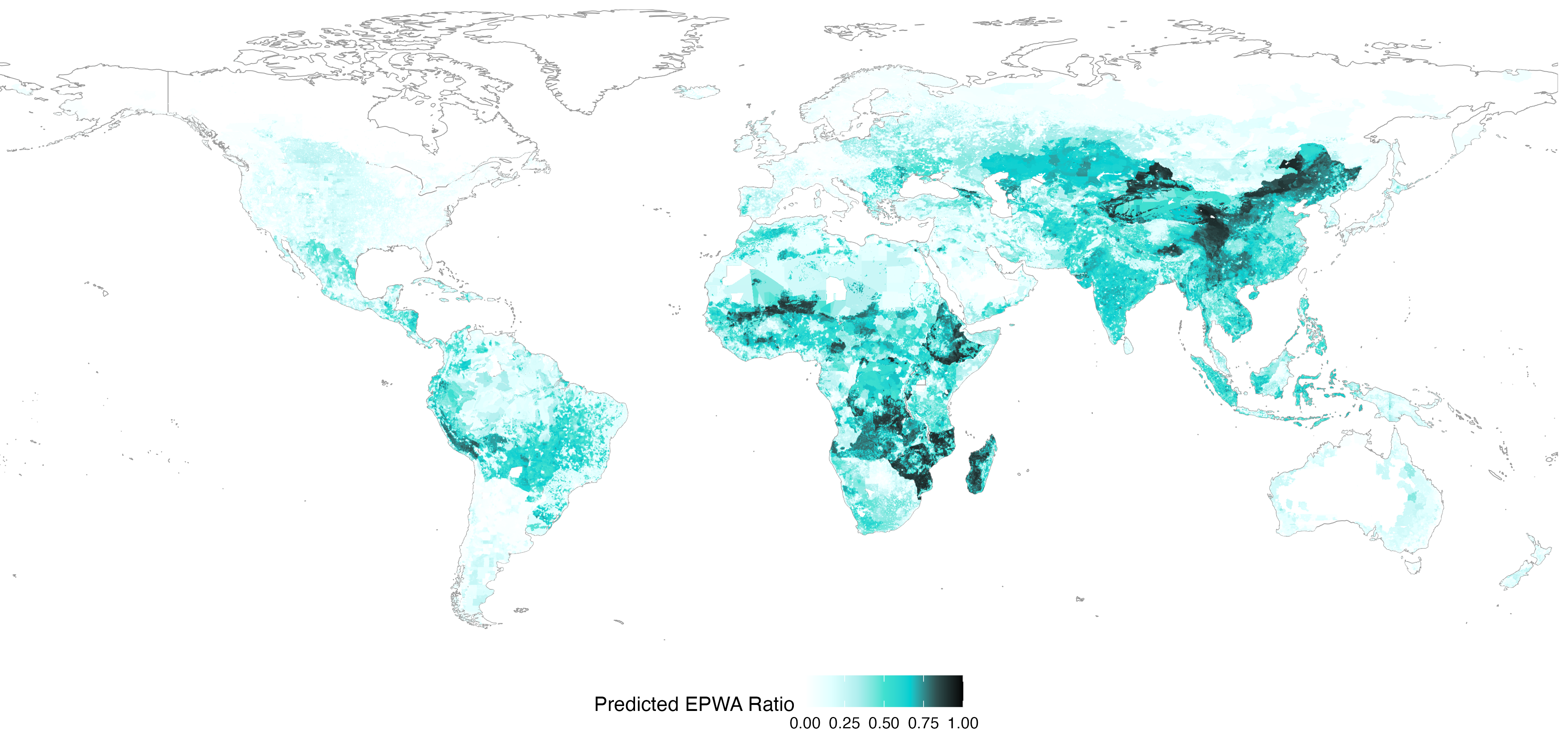}
        \caption{GAMM predictions for 2050.}
        \label{fig:SI_subfig_gam}
    \end{subfigure}
    \caption{Downscaled predictions of the EPWA ratio under SSP2 in year 2050 for the qGAM and GAM models.}
    \label{fig:model_comparison_SI}
\end{figure}

We implemented an equivalent version of the selected GAMM model in a qGAM framework. While the qGAM model is in theory distribution free, we do undertake a logit-transformed ratio of the agricultural workforce as the response, to stabilize the quantile estimates. Figure \ref{fig:rmse_validation} shows the comparison across time and space validation strategies, as well as multiscale validation across regions. Figures \ref{fig:SI_subfig_qgam}, \ref{fig:SI_subfig_gam} show the downscaled ratios of employable population for the equivalent qGAM and GAMM models, respectively. We observe two things from these plots. First, the error of the GAMM model is lower across the majority of validation exercises (although the qGAM does perform better in some regions on the multiscale validation). Second, even if we tried other configurations (e.g., non-transformed response) predictions are clearly out of line with what we would expect from existing literature. 

\subsection{Tree-based approaches}
The three candidate models we explored include XGBoost and LightGBM, and a standard random forest. For each model, all hyperparameters were selected to minimize root mean squared error (RMSE) in 10-fold cross-validation.

\bigskip

\begin{figure}[H]
    \centering
    \includegraphics[width=0.55\textwidth]{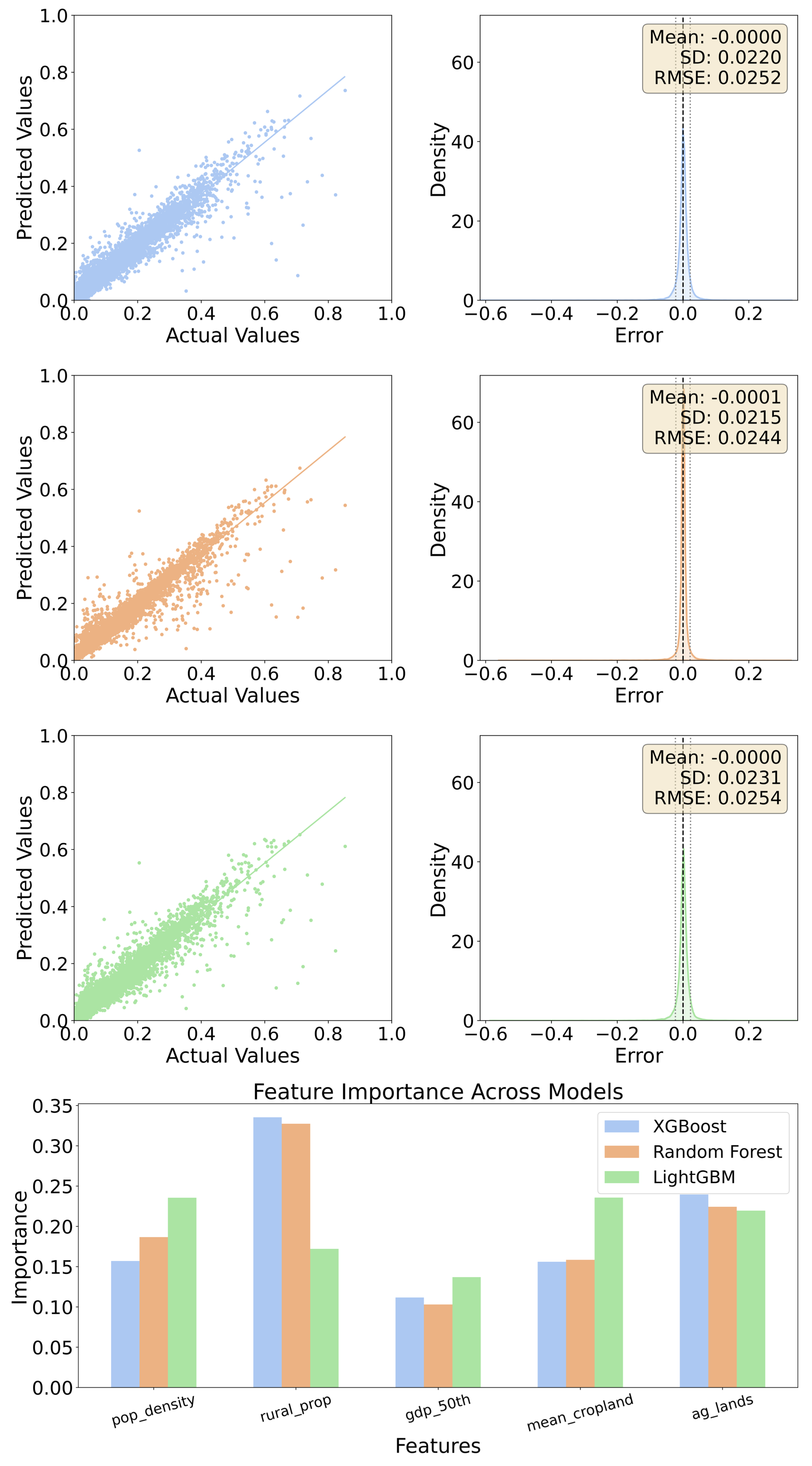}
    \caption{Performance and feature importance plot across tree based model specifications.}
    \label{fig:ml_results}
\end{figure}

Model performance was also evaluated using 10-fold cross-validation. The random forest achieved the lowest cross-validated RMSE of 0.0244 across geographies (see Figure \ref{fig:ml_results}). Errors were evenly distributed around zero, with minimal bias in over- or under-predicting EPWA. Feature importance plots also indicate the critical importance of rural proportion of the population. 

While we found tree based models to perform well in learning the relationships between predictors (like GDP, population density) and EPWA as shown in Figure \ref{fig:ml_results} within the range of the training data, we found these models to generally fail when projecting future values (see Table \ref{tab:ml_projections} for future projections of people working in agriculture by 2050). This inability to forecast is a well-known limitation of tree-based models, which are excellent at capturing complex interactions within the training data but struggle with extrapolation. We also explored these models further through two ensemble approaches: (1) voting with weights selected to minimize RMSE and (2) stacking, which uses the base models’ predictions as predictors in a range of meta-regressors (linear regression, Lasso, Ridge, Random Forest, Gradient Boosting, SVR and a Multi-layer Perceptron regressor). However, the predicted magnitudes and signs appear implausible and inconsistent with theoretical expectations and empirical studies \cite{mehrabi2023likely}. Future research may address this, through use of ensemble approaches with linear predictors or training with data augmentation to improve projections and handle out-of-sample values more effectively.

\begingroup\fontsize{8}{10}\selectfont

\begin{longtable}[t]{llllll}
\caption{\label{tab:ml_projections}Random Forest Uncorrected Predictions of Total Agricultural Population and Percent Change by World Bank Region under scenario SSP2}\\
\toprule
region\_wb & Pop baseline & Pop SSP2 2050 & $\Delta$ Pop SSP2 2050 & Pop SSP2 2100 & $\Delta$ Pop SSP2 2100\\
\midrule
\endfirsthead
\caption[]{Uncorrected Predictions of Total Agricultural Population and Percent Change by World Bank Region under scenario SSP2 \textit{(continued)}}\\
\toprule
region\_wb & Pop baseline & Pop SSP2 2050 & $\Delta$ Pop SSP2 2050 & Pop SSP2 2100 & $\Delta$ Pop SSP2 2100\\
\midrule
\endhead

\endfoot
\bottomrule
\endlastfoot
\cellcolor{gray!10}{EAP} & \cellcolor{gray!10}{152.8M} & \cellcolor{gray!10}{51.2M} & \cellcolor{gray!10}{-66.5\%} & \cellcolor{gray!10}{31.6M} & \cellcolor{gray!10}{-79.3\%}\\
ECA & 20.0M & 14.7M & -26.8\% & 11.1M & -44.5\%\\
\cellcolor{gray!10}{LAC} & \cellcolor{gray!10}{17.7M} & \cellcolor{gray!10}{12.9M} & \cellcolor{gray!10}{-27.2\%} & \cellcolor{gray!10}{8.8M} & \cellcolor{gray!10}{-50.3\%}\\
MENA & 16.7M & 10.3M & -38.1\% & 8.0M & -52.1\%\\
\cellcolor{gray!10}{NA} & \cellcolor{gray!10}{5.3M} & \cellcolor{gray!10}{4.5M} & \cellcolor{gray!10}{-14.7\%} & \cellcolor{gray!10}{3.8M} & \cellcolor{gray!10}{-27.7\%}\\
SA & 211.2M & 60.0M & -71.6\% & 35.9M & -83.0\%\\
\cellcolor{gray!10}{SSA} & \cellcolor{gray!10}{77.1M} & \cellcolor{gray!10}{52.6M} & \cellcolor{gray!10}{-31.8\%} & \cellcolor{gray!10}{44.6M} & \cellcolor{gray!10}{-42.1\%}\\*
\end{longtable}
\endgroup{}

\section{Extended Data}
\begingroup\fontsize{8}{10}\selectfont

\begin{longtable}[t]{llllll}
\caption{\label{tab:ag_pop_ssp1_uncorrected}Uncorrected Predictions of Total Agricultural Population and Percent Change by World Bank Region under scenario SSP1}\\
\toprule
WB Region & Pop baseline & Pop SSP1 2050 & $\Delta$ Pop SSP1 2050 & Pop SSP1 2100 & $\Delta$ Pop SSP1 2100\\
\midrule
\endfirsthead
\caption[]{Uncorrected Predictions of Total Agricultural Population and Percent Change by World Bank Region under scenario SSP1 \textit{(continued)}}\\
\toprule
WB Region & Pop baseline & Pop SSP1 2050 & $\Delta$ Pop SSP1 2050 & Pop SSP1 2100 & $\Delta$ Pop SSP1 2100\\
\midrule
\endhead

\endfoot
\bottomrule
\endlastfoot
\cellcolor{gray!10}{EAP} & \cellcolor{gray!10}{426.8M} & \cellcolor{gray!10}{331.5M} & \cellcolor{gray!10}{-22.3\%} & \cellcolor{gray!10}{202.7M} & \cellcolor{gray!10}{-52.5\%}\\
ECA & 36.5M & 28.0M & -23.4\% & 19.4M & -47.0\%\\
\cellcolor{gray!10}{LAC} & \cellcolor{gray!10}{43.0M} & \cellcolor{gray!10}{38.0M} & \cellcolor{gray!10}{-11.5\%} & \cellcolor{gray!10}{31.3M} & \cellcolor{gray!10}{-27.3\%}\\
MENA & 37.3M & 33.3M & -10.7\% & 29.9M & -19.9\%\\
\cellcolor{gray!10}{NA} & \cellcolor{gray!10}{4.8M} & \cellcolor{gray!10}{4.6M} & \cellcolor{gray!10}{-2.8\%} & \cellcolor{gray!10}{4.1M} & \cellcolor{gray!10}{-13.6\%}\\
SA & 406.4M & 361.4M & -11.1\% & 248.9M & -38.8\%\\
\cellcolor{gray!10}{SSA} & \cellcolor{gray!10}{269.3M} & \cellcolor{gray!10}{307.2M} & \cellcolor{gray!10}{14.1\%} & \cellcolor{gray!10}{264.7M} & \cellcolor{gray!10}{-1.7\%}\\*
\end{longtable}
\endgroup{}

\begingroup\fontsize{8}{10}\selectfont

\begin{longtable}[t]{llllll}
\caption{Corrected Predictions of Total Agricultural Population and Percent Change by World Bank Region under scenario SSP1}\\
\toprule
WB Region & Pop baseline & Pop SSP1 2050 & $\Delta$ Pop SSP1 2050 & Pop SSP1 2100 & $\Delta$ Pop SSP1 2100\\
\midrule
\endfirsthead
\caption[]{Corrected Predictions of Total Agricultural Population and Percent Change by World Bank Region under scenario SSP1 \textit{(continued)}}\\
\toprule
WB Region & Pop baseline & Pop SSP1 2050 & $\Delta$ Pop SSP1 2050 & Pop SSP1 2100 & $\Delta$ Pop SSP1 2100\\
\midrule
\endhead

\endfoot
\bottomrule
\endlastfoot
\cellcolor{gray!10}{EAP} & \cellcolor{gray!10}{422.5M} & \cellcolor{gray!10}{327.5M} & \cellcolor{gray!10}{-22.5\%} & \cellcolor{gray!10}{199.5M} & \cellcolor{gray!10}{-52.8\%}\\
ECA & 34.4M & 26.1M & -24.1\% & 17.5M & -49.2\%\\
\cellcolor{gray!10}{LAC} & \cellcolor{gray!10}{51.2M} & \cellcolor{gray!10}{44.7M} & \cellcolor{gray!10}{-12.8\%} & \cellcolor{gray!10}{37.0M} & \cellcolor{gray!10}{-27.7\%}\\
MENA & 37.8M & 33.7M & -10.8\% & 30.8M & -18.4\%\\
\cellcolor{gray!10}{NA} & \cellcolor{gray!10}{3.9M} & \cellcolor{gray!10}{3.7M} & \cellcolor{gray!10}{-4.5\%} & \cellcolor{gray!10}{3.1M} & \cellcolor{gray!10}{-20.3\%}\\
SA & 402.4M & 358.4M & -10.9\% & 247.7M & -38.4\%\\
\cellcolor{gray!10}{SSA} & \cellcolor{gray!10}{314.8M} & \cellcolor{gray!10}{356.7M} & \cellcolor{gray!10}{13.3\%} & \cellcolor{gray!10}{306.9M} & \cellcolor{gray!10}{-2.5\%}\\*
\end{longtable}
\endgroup{}

\begingroup\fontsize{8}{10}\selectfont

\begin{longtable}[t]{llllll}
\caption{Uncorrected Predictions of Total Agricultural Population and Percent Change by World Bank Region under scenario SSP3}\\
\toprule
WB Region & Pop baseline & Pop SSP3 2050 & $\Delta$ Pop SSP3 2050 & Pop SSP3 2100 & $\Delta$ Pop SSP3 2100\\
\midrule
\endfirsthead
\caption[]{Uncorrected Predictions of Total Agricultural Population and Percent Change by World Bank Region under scenario SSP3 \textit{(continued)}}\\
\toprule
WB Region & Pop baseline & Pop SSP3 2050 & $\Delta$ Pop SSP3 2050 & Pop SSP3 2100 & $\Delta$ Pop SSP3 2100\\
\midrule
\endhead

\endfoot
\bottomrule
\endlastfoot
\cellcolor{gray!10}{EAP} & \cellcolor{gray!10}{426.8M} & \cellcolor{gray!10}{444.7M} & \cellcolor{gray!10}{4.2\%} & \cellcolor{gray!10}{383.1M} & \cellcolor{gray!10}{-10.2\%}\\
ECA & 36.5M & 36.6M & 0.2\% & 33.7M & -7.9\%\\
\cellcolor{gray!10}{LAC} & \cellcolor{gray!10}{43.0M} & \cellcolor{gray!10}{53.2M} & \cellcolor{gray!10}{23.8\%} & \cellcolor{gray!10}{61.0M} & \cellcolor{gray!10}{42.1\%}\\
MENA & 37.3M & 46.5M & 24.6\% & 54.7M & 46.6\%\\
\cellcolor{gray!10}{NA} & \cellcolor{gray!10}{4.8M} & \cellcolor{gray!10}{3.8M} & \cellcolor{gray!10}{-19.5\%} & \cellcolor{gray!10}{2.4M} & \cellcolor{gray!10}{-50.4\%}\\
SA & 406.4M & 558.9M & 37.5\% & 632.2M & 55.6\%\\
\cellcolor{gray!10}{SSA} & \cellcolor{gray!10}{269.3M} & \cellcolor{gray!10}{484.3M} & \cellcolor{gray!10}{79.8\%} & \cellcolor{gray!10}{664.9M} & \cellcolor{gray!10}{146.9\%}\\*
\end{longtable}
\endgroup{}

\begingroup\fontsize{8}{10}\selectfont

\begin{longtable}[t]{llllll}
\caption{Corrected Predictions of Total Agricultural Population and Percent Change by World Bank Region under scenario SSP3}\\
\toprule
WB Region & Pop baseline & Pop SSP3 2050 & $\Delta$ Pop SSP3 2050 & Pop SSP3 2100 & $\Delta$ Pop SSP3 2100\\
\midrule
\endfirsthead
\caption[]{Corrected Predictions of Total Agricultural Population and Percent Change by World Bank Region under scenario SSP3 \textit{(continued)}}\\
\toprule
WB Region & Pop baseline & Pop SSP3 2050 & $\Delta$ Pop SSP3 2050 & Pop SSP3 2100 & $\Delta$ Pop SSP3 2100\\
\midrule
\endhead

\endfoot
\bottomrule
\endlastfoot
\cellcolor{gray!10}{EAP} & \cellcolor{gray!10}{422.5M} & \cellcolor{gray!10}{439.7M} & \cellcolor{gray!10}{4.1\%} & \cellcolor{gray!10}{378.1M} & \cellcolor{gray!10}{-10.5\%}\\
ECA & 34.4M & 34.4M & 0.0\% & 31.9M & -7.4\%\\
\cellcolor{gray!10}{LAC} & \cellcolor{gray!10}{51.2M} & \cellcolor{gray!10}{63.7M} & \cellcolor{gray!10}{24.4\%} & \cellcolor{gray!10}{72.9M} & \cellcolor{gray!10}{42.4\%}\\
MENA & 37.8M & 46.4M & 22.8\% & 54.3M & 43.8\%\\
\cellcolor{gray!10}{NA} & \cellcolor{gray!10}{3.9M} & \cellcolor{gray!10}{3.1M} & \cellcolor{gray!10}{-21.2\%} & \cellcolor{gray!10}{1.8M} & \cellcolor{gray!10}{-53.9\%}\\
\addlinespace
SA & 402.4M & 554.6M & 37.8\% & 630.5M & 56.7\%\\
\cellcolor{gray!10}{SSA} & \cellcolor{gray!10}{314.8M} & \cellcolor{gray!10}{566.6M} & \cellcolor{gray!10}{80.0\%} & \cellcolor{gray!10}{771.7M} & \cellcolor{gray!10}{145.1\%}\\*
\end{longtable}
\endgroup{}

\begingroup\fontsize{8}{10}\selectfont

\begin{longtable}[t]{llllll}
\caption{Uncorrected Predictions of Total Agricultural Population and Percent Change by World Bank Region under scenario SSP4}\\
\toprule
WB Region & Pop baseline & Pop SSP4 2050 & $\Delta$ Pop SSP4 2050 & Pop SSP4 2100 & $\Delta$ Pop SSP4 2100\\
\midrule
\endfirsthead
\caption[]{Uncorrected Predictions of Total Agricultural Population and Percent Change by World Bank Region under scenario SSP4 \textit{(continued)}}\\
\toprule
WB Region & Pop baseline & Pop SSP4 2050 & $\Delta$ Pop SSP4 2050 & Pop SSP4 2100 & $\Delta$ Pop SSP4 2100\\
\midrule
\endhead

\endfoot
\bottomrule
\endlastfoot
\cellcolor{gray!10}{EAP} & \cellcolor{gray!10}{426.8M} & \cellcolor{gray!10}{324.2M} & \cellcolor{gray!10}{-24.0\%} & \cellcolor{gray!10}{177.8M} & \cellcolor{gray!10}{-58.3\%}\\
ECA & 36.5M & 27.8M & -23.8\% & 18.3M & -50.0\%\\
\cellcolor{gray!10}{LAC} & \cellcolor{gray!10}{43.0M} & \cellcolor{gray!10}{39.0M} & \cellcolor{gray!10}{-9.3\%} & \cellcolor{gray!10}{32.6M} & \cellcolor{gray!10}{-24.2\%}\\
MENA & 37.3M & 37.1M & -0.5\% & 40.4M & 8.4\%\\
\cellcolor{gray!10}{NA} & \cellcolor{gray!10}{4.8M} & \cellcolor{gray!10}{4.3M} & \cellcolor{gray!10}{-9.3\%} & \cellcolor{gray!10}{3.3M} & \cellcolor{gray!10}{-30.5\%}\\
SA & 406.4M & 379.4M & -6.6\% & 273.8M & -32.6\%\\
\cellcolor{gray!10}{SSA} & \cellcolor{gray!10}{269.3M} & \cellcolor{gray!10}{377.6M} & \cellcolor{gray!10}{40.2\%} & \cellcolor{gray!10}{458.1M} & \cellcolor{gray!10}{70.1\%}\\*
\end{longtable}
\endgroup{}

\begingroup\fontsize{8}{10}\selectfont

\begin{longtable}[t]{llllll}
\caption{Corrected Predictions of Total Agricultural Population and Percent Change by World Bank Region under scenario SSP4}\\
\toprule
WB Region & Pop baseline & Pop SSP4 2050 & $\Delta$ Pop SSP4 2050 & Pop SSP4 2100 & $\Delta$ Pop SSP4 2100\\
\midrule
\endfirsthead
\caption[]{Corrected Predictions of Total Agricultural Population and Percent Change by World Bank Region under scenario SSP4 \textit{(continued)}}\\
\toprule
WB Region & Pop baseline & Pop SSP4 2050 & $\Delta$ Pop SSP4 2050 & Pop SSP4 2100 & $\Delta$ Pop SSP4 2100\\
\midrule
\endhead

\endfoot
\bottomrule
\endlastfoot
\cellcolor{gray!10}{EAP} & \cellcolor{gray!10}{422.5M} & \cellcolor{gray!10}{320.3M} & \cellcolor{gray!10}{-24.2\%} & \cellcolor{gray!10}{174.8M} & \cellcolor{gray!10}{-58.6\%}\\
ECA & 34.4M & 26.0M & -24.4\% & 16.7M & -51.4\%\\
\cellcolor{gray!10}{LAC} & \cellcolor{gray!10}{51.2M} & \cellcolor{gray!10}{45.7M} & \cellcolor{gray!10}{-10.8\%} & \cellcolor{gray!10}{38.7M} & \cellcolor{gray!10}{-24.5\%}\\
MENA & 37.8M & 37.3M & -1.2\% & 40.9M & 8.4\%\\
\cellcolor{gray!10}{NA} & \cellcolor{gray!10}{3.9M} & \cellcolor{gray!10}{3.5M} & \cellcolor{gray!10}{-11.0\%} & \cellcolor{gray!10}{2.5M} & \cellcolor{gray!10}{-36.0\%}\\
SA & 402.4M & 376.4M & -6.5\% & 273.8M & -32.0\%\\
\cellcolor{gray!10}{SSA} & \cellcolor{gray!10}{314.8M} & \cellcolor{gray!10}{440.1M} & \cellcolor{gray!10}{39.8\%} & \cellcolor{gray!10}{533.6M} & \cellcolor{gray!10}{69.5\%}\\*
\end{longtable}
\endgroup{}

\begingroup\fontsize{8}{10}\selectfont

\begin{longtable}[t]{llllll}
\caption{Uncorrected Predictions of Total Agricultural Population and Percent Change by World Bank Region under scenario SSP5}\\
\toprule
WB Region & Pop baseline & Pop SSP5 2050 & $\Delta$ Pop SSP5 2050 & Pop SSP5 2100 & $\Delta$ Pop SSP5 2100\\
\midrule
\endfirsthead
\caption[]{Uncorrected Predictions of Total Agricultural Population and Percent Change by World Bank Region under scenario SSP5 \textit{(continued)}}\\
\toprule
WB Region & Pop baseline & Pop SSP5 2050 & $\Delta$ Pop SSP5 2050 & Pop SSP5 2100 & $\Delta$ Pop SSP5 2100\\
\midrule
\endhead

\endfoot
\bottomrule
\endlastfoot
\cellcolor{gray!10}{EAP} & \cellcolor{gray!10}{426.8M} & \cellcolor{gray!10}{330.5M} & \cellcolor{gray!10}{-22.6\%} & \cellcolor{gray!10}{207.1M} & \cellcolor{gray!10}{-51.5\%}\\
ECA & 36.5M & 29.0M & -20.7\% & 24.7M & -32.5\%\\
\cellcolor{gray!10}{LAC} & \cellcolor{gray!10}{43.0M} & \cellcolor{gray!10}{37.8M} & \cellcolor{gray!10}{-12.1\%} & \cellcolor{gray!10}{33.2M} & \cellcolor{gray!10}{-22.8\%}\\
MENA & 37.3M & 33.4M & -10.5\% & 32.8M & -12.1\%\\
\cellcolor{gray!10}{NA} & \cellcolor{gray!10}{4.8M} & \cellcolor{gray!10}{5.3M} & \cellcolor{gray!10}{11.2\%} & \cellcolor{gray!10}{6.8M} & \cellcolor{gray!10}{42.1\%}\\
SA & 406.4M & 355.9M & -12.4\% & 238.7M & -41.3\%\\
\cellcolor{gray!10}{SSA} & \cellcolor{gray!10}{269.3M} & \cellcolor{gray!10}{297.3M} & \cellcolor{gray!10}{10.4\%} & \cellcolor{gray!10}{250.1M} & \cellcolor{gray!10}{-7.1\%}\\*
\end{longtable}
\endgroup{}

\begingroup\fontsize{8}{10}\selectfont

\begin{longtable}[t]{llllll}
\caption{\label{tab:ag_pop_ssp5_corrected}Corrected Predictions of Total Agricultural Population and Percent Change by World Bank Region under scenario SSP5}\\
\toprule
WB Region & Pop baseline & Pop SSP5 2050 & $\Delta$ Pop SSP5 2050 & Pop SSP5 2100 & $\Delta$ Pop SSP5 2100\\
\midrule
\endfirsthead
\caption[]{Corrected Predictions of Total Agricultural Population and Percent Change by World Bank Region under scenario SSP5 \textit{(continued)}}\\
\toprule
WB Region & Pop baseline & Pop SSP5 2050 & $\Delta$ Pop SSP5 2050 & Pop SSP5 2100 & $\Delta$ Pop SSP5 2100\\
\midrule
\endhead

\endfoot
\bottomrule
\endlastfoot
\cellcolor{gray!10}{EAP} & \cellcolor{gray!10}{422.5M} & \cellcolor{gray!10}{326.3M} & \cellcolor{gray!10}{-22.8\%} & \cellcolor{gray!10}{202.9M} & \cellcolor{gray!10}{-52.0\%}\\
ECA & 34.4M & 26.9M & -21.9\% & 20.8M & -39.5\%\\
\cellcolor{gray!10}{LAC} & \cellcolor{gray!10}{51.2M} & \cellcolor{gray!10}{44.3M} & \cellcolor{gray!10}{-13.5\%} & \cellcolor{gray!10}{38.9M} & \cellcolor{gray!10}{-24.1\%}\\
MENA & 37.8M & 33.8M & -10.5\% & 33.6M & -10.9\%\\
\cellcolor{gray!10}{NA} & \cellcolor{gray!10}{3.9M} & \cellcolor{gray!10}{4.3M} & \cellcolor{gray!10}{9.2\%} & \cellcolor{gray!10}{4.7M} & \cellcolor{gray!10}{21.4\%}\\
\addlinespace
SA & 402.4M & 353.3M & -12.2\% & 237.5M & -41.0\%\\
\cellcolor{gray!10}{SSA} & \cellcolor{gray!10}{314.8M} & \cellcolor{gray!10}{345.0M} & \cellcolor{gray!10}{9.6\%} & \cellcolor{gray!10}{289.0M} & \cellcolor{gray!10}{-8.2\%}\\*
\end{longtable}
\endgroup{}

\begingroup\fontsize{8}{10}\selectfont

\begin{longtable}[t]{l|>{}r|>{}r|>{}r|>{}r|>{}r}
\caption{\label{tab:top_ag_changes2100} Top Countries by Largest Positive and Negative Absolute Changes in agricultural population by 2100 under SSP 2. Population in millions (M).}\\
\toprule
Country & WB Region & Type & Baseline Pop (M) & Future Pop (M) & Pop Change (M) \\
\midrule
\endfirsthead
\caption[]{Top Countries by Largest Positive and Negative Absolute Changes in agricultural population by 2100 under SSP 2. Population in millions (M). \textit{(continued)}}\\
\toprule
Country & WB Region & Type & Baseline Pop (M) & Future Pop (M) & Pop Change (M) \\
\midrule
\endhead

\endfoot
\bottomrule
\endlastfoot
\cellcolor{gray!10}{Nigeria} & \cellcolor{gray!10}{SSA} & \cellcolor{gray!10}{Largest Positive Change} & \cellcolor{gray!10}{39.68} & \cellcolor{gray!10}{73.29} & \cellcolor{gray!10}{33.61}\\
Uganda & SSA & Largest Positive Change & 18.04 & 30.77 & 12.72\\
\cellcolor{gray!10}{Dem. Republic of the Congo} & \cellcolor{gray!10}{SSA} & \cellcolor{gray!10}{Largest Positive Change} & \cellcolor{gray!10}{27.90} & \cellcolor{gray!10}{36.75} & \cellcolor{gray!10}{8.85}\\
Philippines & EAP & Largest Positive Change & 14.40 & 22.05 & 7.66\\
\cellcolor{gray!10}{Niger} & \cellcolor{gray!10}{SSA} & \cellcolor{gray!10}{Largest Positive Change} & \cellcolor{gray!10}{7.69} & \cellcolor{gray!10}{13.89} & \cellcolor{gray!10}{6.19}\\
\addlinespace
Malawi & SSA & Largest Positive Change & 6.26 & 11.93 & 5.68\\
\cellcolor{gray!10}{Angola} & \cellcolor{gray!10}{SSA} & \cellcolor{gray!10}{Largest Positive Change} & \cellcolor{gray!10}{5.32} & \cellcolor{gray!10}{9.50} & \cellcolor{gray!10}{4.18}\\
Mozambique & SSA & Largest Positive Change & 13.01 & 16.93 & 3.92\\
\cellcolor{gray!10}{Zambia} & \cellcolor{gray!10}{SSA} & \cellcolor{gray!10}{Largest Positive Change} & \cellcolor{gray!10}{4.29} & \cellcolor{gray!10}{7.81} & \cellcolor{gray!10}{3.52}\\
Ghana & SSA & Largest Positive Change & 6.45 & 9.46 & 3.01\\
\addlinespace
\cellcolor{gray!10}{Egypt} & \cellcolor{gray!10}{MENA} & \cellcolor{gray!10}{Largest Positive Change} & \cellcolor{gray!10}{6.02} & \cellcolor{gray!10}{8.43} & \cellcolor{gray!10}{2.41}\\
Afghanistan & SA & Largest Positive Change & 7.16 & 9.45 & 2.29\\
\cellcolor{gray!10}{Mali} & \cellcolor{gray!10}{SSA} & \cellcolor{gray!10}{Largest Positive Change} & \cellcolor{gray!10}{7.69} & \cellcolor{gray!10}{9.96} & \cellcolor{gray!10}{2.27}\\
Guatemala & LAC & Largest Positive Change & 2.86 & 4.71 & 1.85\\
\cellcolor{gray!10}{Rwanda} & \cellcolor{gray!10}{SSA} & \cellcolor{gray!10}{Largest Positive Change} & \cellcolor{gray!10}{3.70} & \cellcolor{gray!10}{5.42} & \cellcolor{gray!10}{1.72}\\
\addlinespace
Madagascar & SSA & Largest Positive Change & 10.72 & 12.41 & 1.69\\
\cellcolor{gray!10}{Burundi} & \cellcolor{gray!10}{SSA} & \cellcolor{gray!10}{Largest Positive Change} & \cellcolor{gray!10}{6.65} & \cellcolor{gray!10}{8.24} & \cellcolor{gray!10}{1.59}\\
Kenya & SSA & Largest Positive Change & 6.06 & 7.63 & 1.57\\
\cellcolor{gray!10}{Pakistan} & \cellcolor{gray!10}{SA} & \cellcolor{gray!10}{Largest Positive Change} & \cellcolor{gray!10}{35.28} & \cellcolor{gray!10}{36.74} & \cellcolor{gray!10}{1.46}\\
Malaysia & EAP & Largest Positive Change & 2.17 & 3.59 & 1.42\\
\addlinespace
\cellcolor{gray!10}{Peru} & \cellcolor{gray!10}{LAC} & \cellcolor{gray!10}{Largest Negative Change} & \cellcolor{gray!10}{3.89} & \cellcolor{gray!10}{2.75} & \cellcolor{gray!10}{-1.13}\\
Romania & ECA & Largest Negative Change & 2.53 & 1.37 & -1.16\\
\cellcolor{gray!10}{Poland} & \cellcolor{gray!10}{ECA} & \cellcolor{gray!10}{Largest Negative Change} & \cellcolor{gray!10}{2.49} & \cellcolor{gray!10}{1.33} & \cellcolor{gray!10}{-1.16}\\
Brazil & LAC & Largest Negative Change & 12.67 & 11.46 & -1.20\\
\cellcolor{gray!10}{Guinea} & \cellcolor{gray!10}{SSA} & \cellcolor{gray!10}{Largest Negative Change} & \cellcolor{gray!10}{3.16} & \cellcolor{gray!10}{1.69} & \cellcolor{gray!10}{-1.46}\\
\addlinespace
Cambodia & EAP & Largest Negative Change & 5.30 & 3.66 & -1.64\\
\cellcolor{gray!10}{Zimbabwe} & \cellcolor{gray!10}{SSA} & \cellcolor{gray!10}{Largest Negative Change} & \cellcolor{gray!10}{4.82} & \cellcolor{gray!10}{3.13} & \cellcolor{gray!10}{-1.69}\\
Japan & EAP & Largest Negative Change & 3.20 & 1.44 & -1.76\\
\cellcolor{gray!10}{Sri Lanka} & \cellcolor{gray!10}{SA} & \cellcolor{gray!10}{Largest Negative Change} & \cellcolor{gray!10}{2.98} & \cellcolor{gray!10}{1.12} & \cellcolor{gray!10}{-1.86}\\
Türkiye & MENA & Largest Negative Change & 7.47 & 5.56 & -1.91\\
\addlinespace
\cellcolor{gray!10}{Russian Federation} & \cellcolor{gray!10}{ECA} & \cellcolor{gray!10}{Largest Negative Change} & \cellcolor{gray!10}{5.91} & \cellcolor{gray!10}{3.49} & \cellcolor{gray!10}{-2.41}\\
Tanzania, United Republic of & SSA & Largest Negative Change & 15.40 & 12.49 & -2.91\\
\cellcolor{gray!10}{Nepal} & \cellcolor{gray!10}{SA} & \cellcolor{gray!10}{Largest Negative Change} & \cellcolor{gray!10}{10.29} & \cellcolor{gray!10}{7.09} & \cellcolor{gray!10}{-3.20}\\
Viet Nam & EAP & Largest Negative Change & 25.74 & 22.36 & -3.37\\
\cellcolor{gray!10}{Myanmar} & \cellcolor{gray!10}{EAP} & \cellcolor{gray!10}{Largest Negative Change} & \cellcolor{gray!10}{12.67} & \cellcolor{gray!10}{7.85} & \cellcolor{gray!10}{-4.82}\\
\addlinespace
Thailand & EAP & Largest Negative Change & 15.81 & 10.10 & -5.71\\
\cellcolor{gray!10}{Indonesia} & \cellcolor{gray!10}{EAP} & \cellcolor{gray!10}{Largest Negative Change} & \cellcolor{gray!10}{43.32} & \cellcolor{gray!10}{33.63} & \cellcolor{gray!10}{-9.69}\\
Bangladesh & SA & Largest Negative Change & 37.65 & 27.75 & -9.90\\
\cellcolor{gray!10}{India} & \cellcolor{gray!10}{SA} & \cellcolor{gray!10}{Largest Negative Change} & \cellcolor{gray!10}{307.86} & \cellcolor{gray!10}{263.57} & \cellcolor{gray!10}{-44.29}\\
China & EAP & Largest Negative Change & 295.66 & 142.53 & -153.14\\*
\end{longtable}
\endgroup{}

\begingroup\fontsize{8}{10}\selectfont

\begin{longtable}[t]{l|>{}r|>{}r|>{}r|>{}r|>{}r}
\caption{\label{tab:top_ag_changes2050} Top Countries by Largest Positive and Negative Absolute Changes in agricultural population by 2050 under SSP 2. Population in millions (M).}\\
\toprule
Country & WB Region & Type & Baseline Pop (M) & Future Pop (M) & Pop Change (M) \\
\midrule
\endfirsthead
\caption[]{Top Countries by Largest Positive and Negative Absolute Changes in agricultural population by 2050 under SSP 2. Population in millions (M). \textit{(continued)}}\\
\toprule
Country & WB Region & Type & Baseline Pop (M) & Future Pop (M) & Pop Change (M) \\
\midrule
\endhead

\endfoot
\bottomrule
\endlastfoot
\cellcolor{gray!10}{India} & \cellcolor{gray!10}{SA} & \cellcolor{gray!10}{Largest Positive Change} & \cellcolor{gray!10}{307.86} & \cellcolor{gray!10}{334.98} & \cellcolor{gray!10}{27.12}\\
Nigeria & SSA & Largest Positive Change & 39.68 & 59.14 & 19.46\\
\cellcolor{gray!10}{Ethiopia} & \cellcolor{gray!10}{SSA} & \cellcolor{gray!10}{Largest Positive Change} & \cellcolor{gray!10}{44.83} & \cellcolor{gray!10}{63.26} & \cellcolor{gray!10}{18.43}\\
Uganda & SSA & Largest Positive Change & 18.04 & 31.21 & 13.16\\
\cellcolor{gray!10}{Dem. Republic of the Congo} & \cellcolor{gray!10}{SSA} & \cellcolor{gray!10}{Largest Positive Change} & \cellcolor{gray!10}{27.90} & \cellcolor{gray!10}{38.73} & \cellcolor{gray!10}{10.83}\\
\addlinespace
Niger & SSA & Largest Positive Change & 7.69 & 15.73 & 8.04\\
\cellcolor{gray!10}{Tanzania, United Republic of} & \cellcolor{gray!10}{SSA} & \cellcolor{gray!10}{Largest Positive Change} & \cellcolor{gray!10}{15.40} & \cellcolor{gray!10}{21.66} & \cellcolor{gray!10}{6.26}\\
Pakistan & SA & Largest Positive Change & 35.28 & 41.18 & 5.90\\
\cellcolor{gray!10}{Madagascar} & \cellcolor{gray!10}{SSA} & \cellcolor{gray!10}{Largest Positive Change} & \cellcolor{gray!10}{10.72} & \cellcolor{gray!10}{15.67} & \cellcolor{gray!10}{4.95}\\
Malawi & SSA & Largest Positive Change & 6.26 & 10.98 & 4.72\\
\addlinespace
\cellcolor{gray!10}{Mali} & \cellcolor{gray!10}{SSA} & \cellcolor{gray!10}{Largest Positive Change} & \cellcolor{gray!10}{7.69} & \cellcolor{gray!10}{11.84} & \cellcolor{gray!10}{4.15}\\
Mozambique & SSA & Largest Positive Change & 13.01 & 17.05 & 4.05\\
\cellcolor{gray!10}{Afghanistan} & \cellcolor{gray!10}{SA} & \cellcolor{gray!10}{Largest Positive Change} & \cellcolor{gray!10}{7.16} & \cellcolor{gray!10}{10.60} & \cellcolor{gray!10}{3.44}\\
Burundi & SSA & Largest Positive Change & 6.65 & 9.78 & 3.14\\
\cellcolor{gray!10}{Angola} & \cellcolor{gray!10}{SSA} & \cellcolor{gray!10}{Largest Positive Change} & \cellcolor{gray!10}{5.32} & \cellcolor{gray!10}{8.17} & \cellcolor{gray!10}{2.84}\\
\addlinespace
Zambia & SSA & Largest Positive Change & 4.29 & 6.91 & 2.62\\
\cellcolor{gray!10}{Philippines} & \cellcolor{gray!10}{EAP} & \cellcolor{gray!10}{Largest Positive Change} & \cellcolor{gray!10}{14.40} & \cellcolor{gray!10}{16.75} & \cellcolor{gray!10}{2.35}\\
Nepal & SA & Largest Positive Change & 10.29 & 12.44 & 2.15\\
\cellcolor{gray!10}{Cameroon} & \cellcolor{gray!10}{SSA} & \cellcolor{gray!10}{Largest Positive Change} & \cellcolor{gray!10}{5.65} & \cellcolor{gray!10}{7.80} & \cellcolor{gray!10}{2.15}\\
Ghana & SSA & Largest Positive Change & 6.45 & 8.50 & 2.05\\
\addlinespace
\cellcolor{gray!10}{Republic of Moldova} & \cellcolor{gray!10}{ECA} & \cellcolor{gray!10}{Largest Negative Change} & \cellcolor{gray!10}{0.65} & \cellcolor{gray!10}{0.43} & \cellcolor{gray!10}{-0.22}\\
United States & NA & Largest Negative Change & 4.20 & 3.98 & -0.23\\
\cellcolor{gray!10}{Serbia} & \cellcolor{gray!10}{ECA} & \cellcolor{gray!10}{Largest Negative Change} & \cellcolor{gray!10}{0.82} & \cellcolor{gray!10}{0.57} & \cellcolor{gray!10}{-0.26}\\
Peru & LAC & Largest Negative Change & 3.89 & 3.47 & -0.42\\
\cellcolor{gray!10}{Egypt} & \cellcolor{gray!10}{MENA} & \cellcolor{gray!10}{Largest Negative Change} & \cellcolor{gray!10}{6.02} & \cellcolor{gray!10}{5.59} & \cellcolor{gray!10}{-0.43}\\
\addlinespace
Ukraine & ECA & Largest Negative Change & 3.01 & 2.57 & -0.43\\
\cellcolor{gray!10}{Romania} & \cellcolor{gray!10}{ECA} & \cellcolor{gray!10}{Largest Negative Change} & \cellcolor{gray!10}{2.53} & \cellcolor{gray!10}{2.05} & \cellcolor{gray!10}{-0.48}\\
Zimbabwe & SSA & Largest Negative Change & 4.82 & 4.30 & -0.52\\
\cellcolor{gray!10}{Dem. People's Rep. of Korea} & \cellcolor{gray!10}{EAP} & \cellcolor{gray!10}{Largest Negative Change} & \cellcolor{gray!10}{1.68} & \cellcolor{gray!10}{1.16} & \cellcolor{gray!10}{-0.52}\\
Türkiye & MENA & Largest Negative Change & 7.47 & 6.91 & -0.55\\
\addlinespace
\cellcolor{gray!10}{Morocco} & \cellcolor{gray!10}{MENA} & \cellcolor{gray!10}{Largest Negative Change} & \cellcolor{gray!10}{3.44} & \cellcolor{gray!10}{2.88} & \cellcolor{gray!10}{-0.56}\\
Poland & ECA & Largest Negative Change & 2.49 & 1.93 & -0.56\\
\cellcolor{gray!10}{Japan} & \cellcolor{gray!10}{EAP} & \cellcolor{gray!10}{Largest Negative Change} & \cellcolor{gray!10}{3.20} & \cellcolor{gray!10}{2.21} & \cellcolor{gray!10}{-0.99}\\
Russian Federation & ECA & Largest Negative Change & 5.91 & 4.72 & -1.19\\
\cellcolor{gray!10}{Thailand} & \cellcolor{gray!10}{EAP} & \cellcolor{gray!10}{Largest Negative Change} & \cellcolor{gray!10}{15.81} & \cellcolor{gray!10}{14.53} & \cellcolor{gray!10}{-1.28}\\
Myanmar & EAP & Largest Negative Change & 12.67 & 11.37 & -1.30\\
\cellcolor{gray!10}{Sri Lanka} & \cellcolor{gray!10}{SA} & \cellcolor{gray!10}{Largest Negative Change} & \cellcolor{gray!10}{2.98} & \cellcolor{gray!10}{1.66} & \cellcolor{gray!10}{-1.32}\\
Bangladesh & SA & Largest Negative Change & 37.65 & 36.18 & -1.48\\
\cellcolor{gray!10}{Indonesia} & \cellcolor{gray!10}{EAP} & \cellcolor{gray!10}{Largest Negative Change} & \cellcolor{gray!10}{43.32} & \cellcolor{gray!10}{39.35} & \cellcolor{gray!10}{-3.96}\\
China & EAP & Largest Negative Change & 295.66 & 254.12 & -41.54\\*
\end{longtable}
\endgroup{}

\begin{figure}[H]
    \centering
    \begin{subfigure}[t]{\textwidth}
        \centering
        \includegraphics[width=0.6\textwidth]{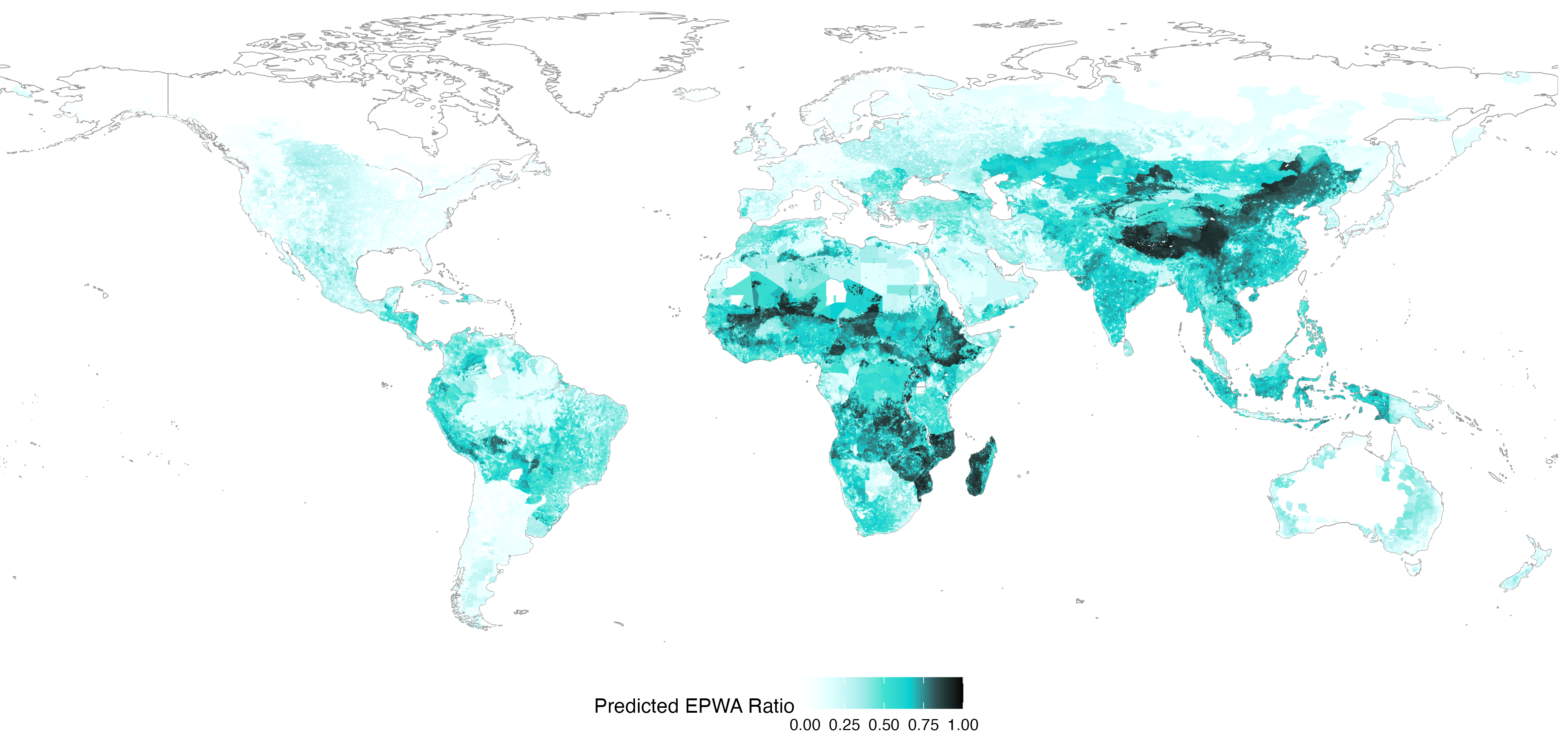}
        \caption{GAM predictions for 2000.}
        \label{fig:subfig_gam2000}
    \end{subfigure}
    \vspace{0.5cm} 
    \begin{subfigure}[t]{\textwidth}
        \centering
        \includegraphics[width=0.6\textwidth]{Figures/ag_epwa_ratio_2_2000_2050.png}
        \caption{GAM predictions for 2050.}
        \label{fig:subfig_gam2050}
    \end{subfigure}
    \vspace{0.1cm} 
    \begin{subfigure}[t]{\textwidth}
        \centering
        \includegraphics[width=0.6\textwidth]{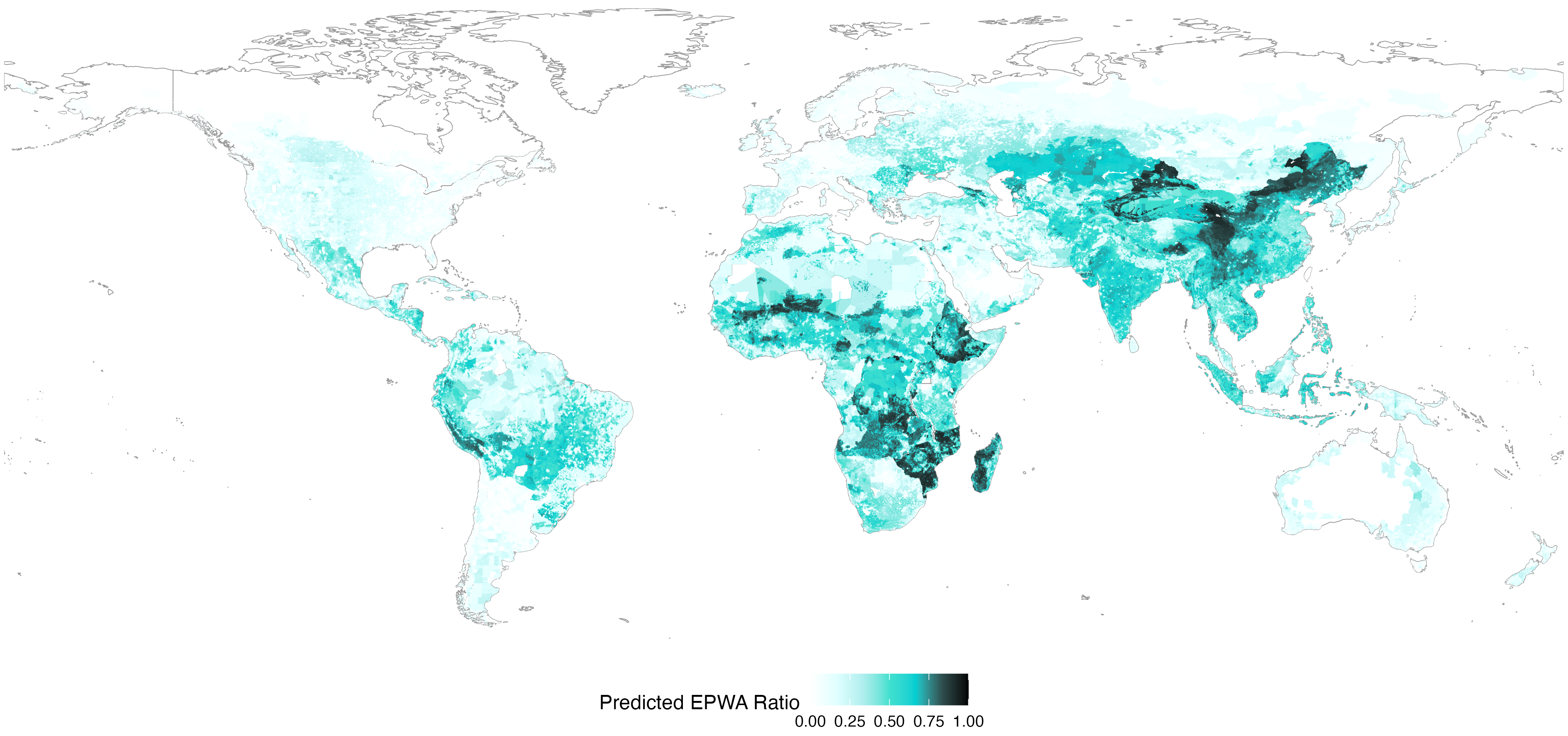}
        \caption{GAM predictions for 2100.}
        \label{fig:subfig_gam2100}
    \end{subfigure}
    \caption{Downscaled predictions of the EPWA ratio under business as usual scenario in years 2000, 2050, 2100 for SSP2}
    \label{fig:mainfig_ratios}
\end{figure}

\begin{figure}[H]
    \centering
    \includegraphics[width=\textwidth]{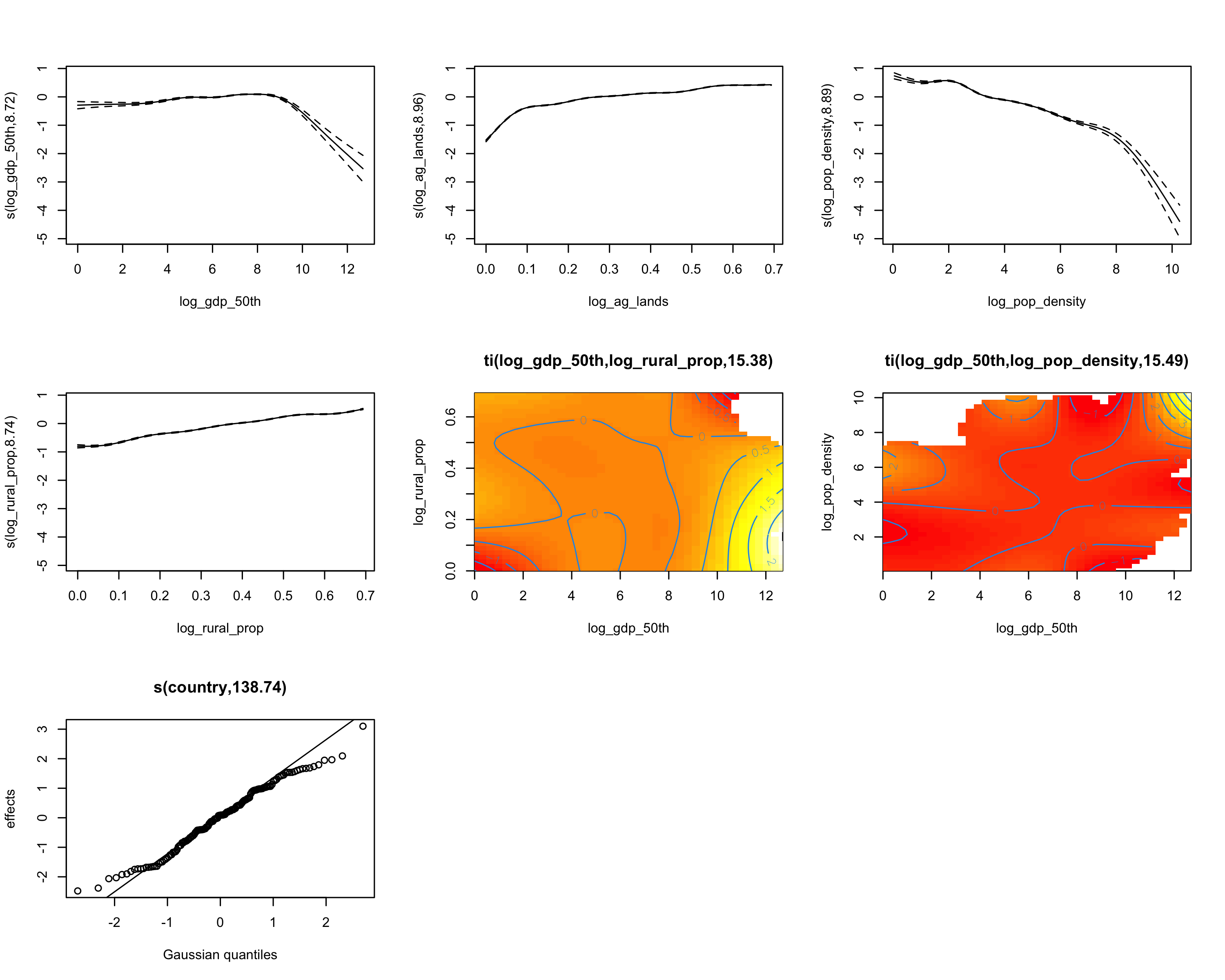}
    \caption{Smooth functions from the selected GAMM Beta regression family with country-level random effects and interactions of population density and rural proportion with GPD. First, the agricultural workforce ratio is overall constant with increasing GDP per capita, showing a sharp decline after a certain threshold, reflecting economic diversification in wealthier regions. Agricultural workforce increases sharply with increases in lower proportions of agricultural land but the growth stabilizes as land availability grows, indicating diminishing returns in workforce absorption. Population density has a strong negative effect, with areas of higher density showing steeper declines in the agricultural workforce, suggesting urbanization's role in reducing agricultural employment. Rural population proportion positively influences the workforce ratio in a relatively linear fashion, reflecting the natural correlation between rurality and agricultural employment.}
    \label{fig:smooth_terms}
\end{figure}

\begin{figure}[H]
    \centering
    \begin{subfigure}[b]{0.6\textwidth}
        \includegraphics[width=\textwidth]{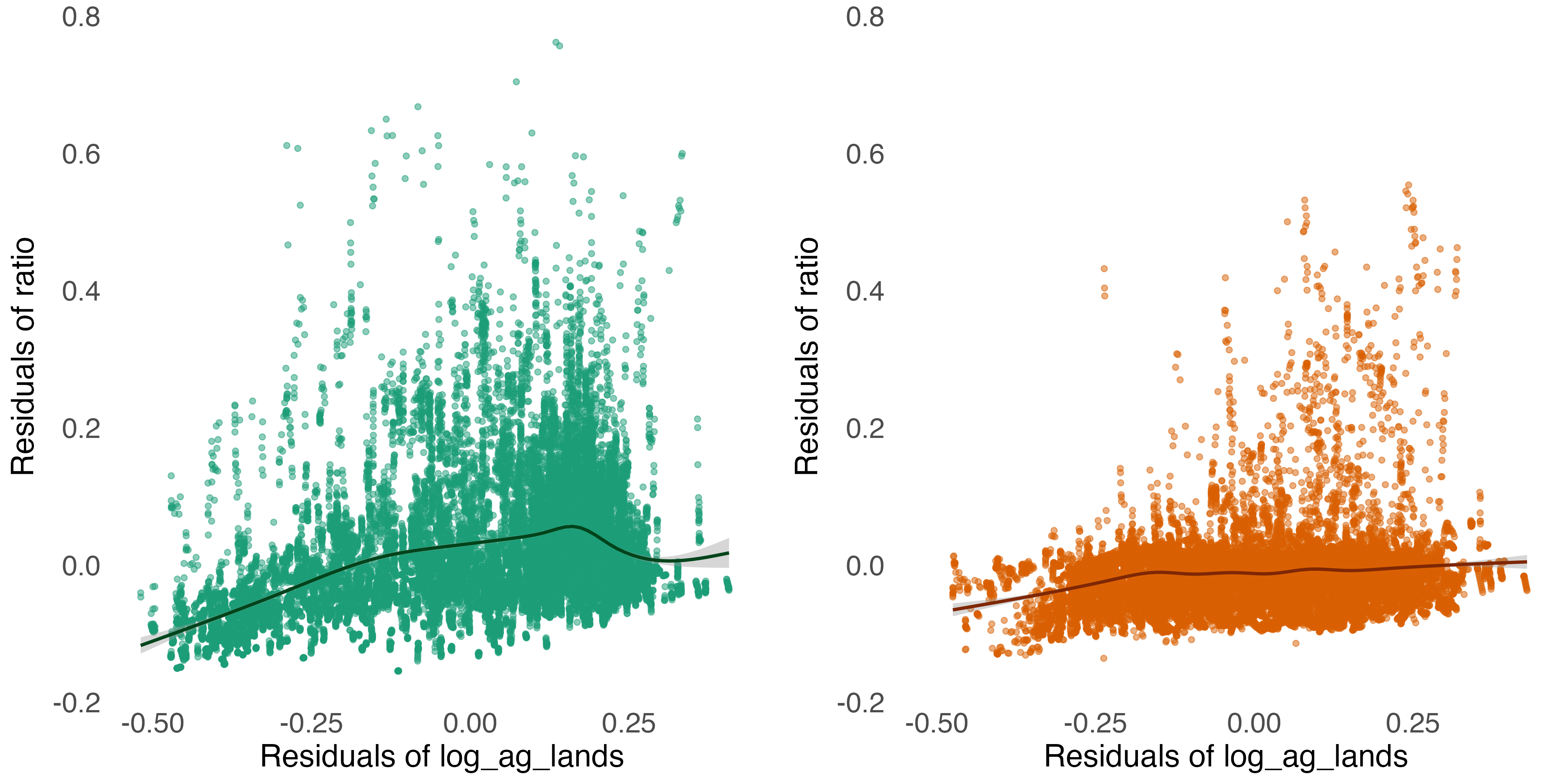}
        \caption{Residuals of Ratio vs. Residuals of Log Agricultural Lands}
        \label{fig:log_ag_lands}
    \end{subfigure}
    
    \begin{subfigure}[b]{0.6\textwidth}
        \includegraphics[width=\textwidth]{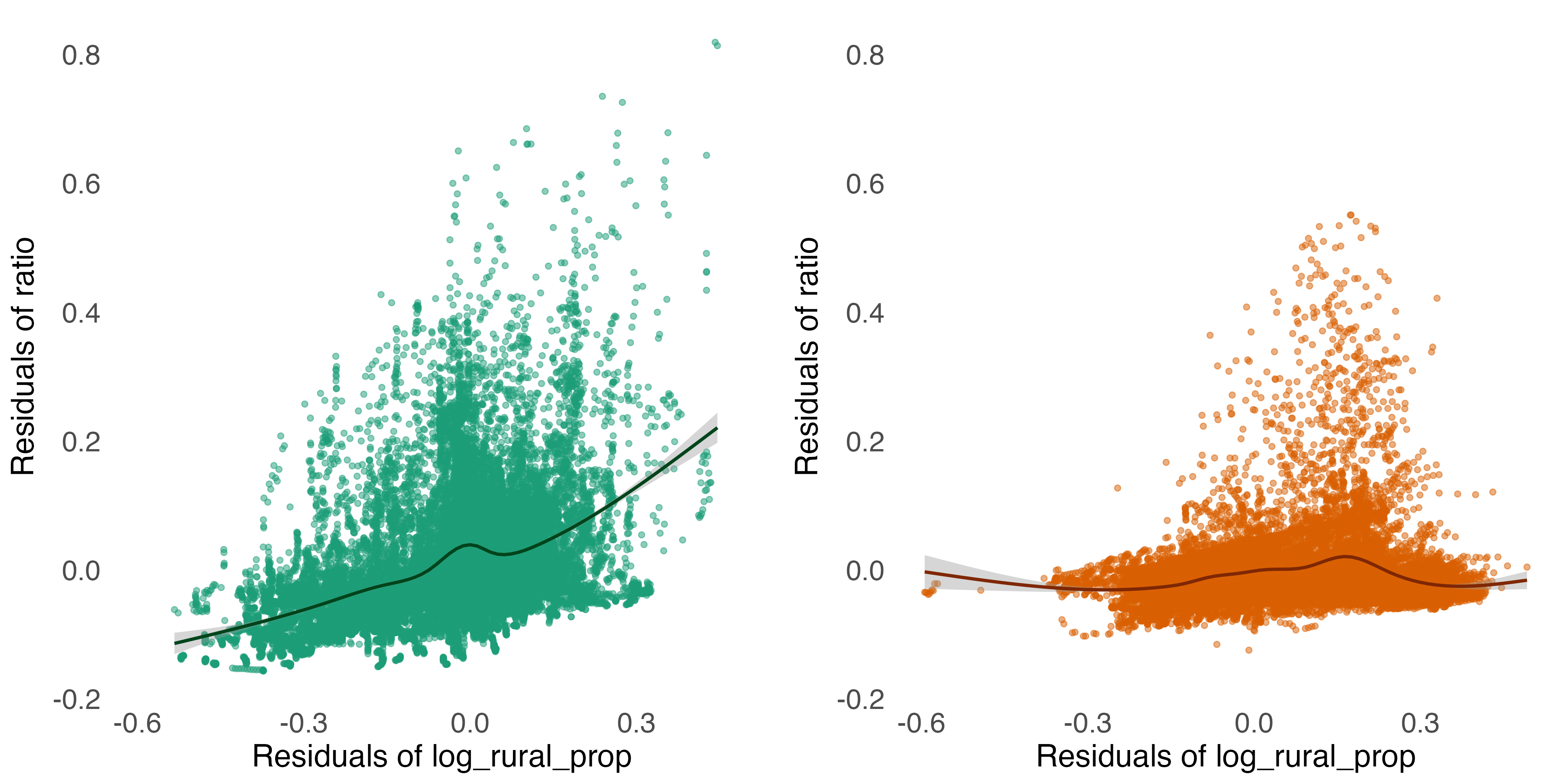}
        \caption{Residuals of Ratio vs. Residuals of Log Rural Proportion}
        \label{fig:log_pop_density}
    \end{subfigure}

    \begin{subfigure}[b]{0.6\textwidth}
        \includegraphics[width=\textwidth]{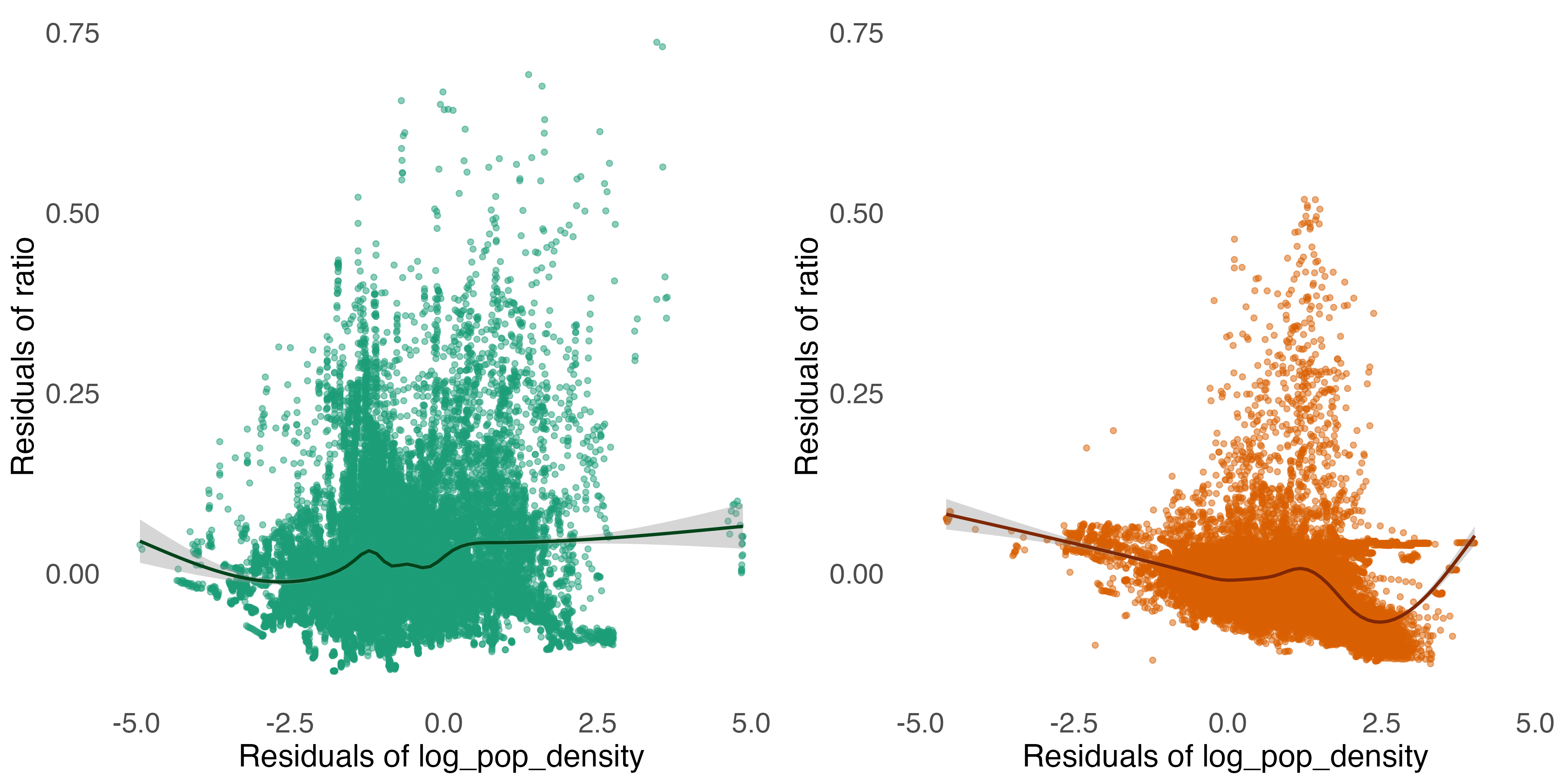}
        \caption{Residuals of Ratio vs. Residuals of Log Population Density}
        \label{fig:log_rural_prop}
    \end{subfigure}
    \caption{Residual plots showing the relationships between the outcome variable (ratio of agricultural population over total population) and different predictors (log-transformed agricultural lands, population density, and rural proportion) for low and high levels of GDP. Residuals were obtained from a generalized additive model (GAMM) applied to isolate the effects of each predictor after adjusting for the remaining variables. The smooth lines in the plots were also generated using GAMM to better capture potential non-linear relationships. These plots show the necessity of including interaction terms in the model to capture the varying influences of key predictors across different economic contexts.}
    \label{fig:residual_relationships}
\end{figure}

\end{document}